\documentclass[%
 reprint,
 nofootinbib,
 amsmath,amssymb,
 aps,
 prx,
]{revtex4-2}

\usepackage[mode=buildnew]{standalone}
\usepackage{amsmath}
\usepackage{graphicx}
\usepackage{dcolumn}
\usepackage{bm}
\usepackage{braket}
\usepackage{subcaption}
\usepackage{float}
\usepackage{color} 

\usepackage{xcolor} 

\usepackage[hidelinks]{hyperref}
\def\equationautorefname#1#2\null{%
  Eq.#1(#2\null)%
}
\def\tableautorefname#1#2\null{%
  {table#1#2\null}%
}
\def\sectionautorefname#1#2\null{%
  {section#1#2\null}%
}
\def\figureautorefname#1#2\null{%
  {Fig.#1#2\null}%
}
\def\subsectionautorefname#1#2\null{%
  {section#1#2\null}%
}

\newcommand{\add}[1]{\textcolor{black}{#1}}

\begin{document}

\captionsetup{justification=raggedright}

\preprint{APS/123-QED}

\title{
Thermodynamic inference in partially accessible Markov networks:\\
A unifying perspective from transition-based waiting time distributions}

\author{Jann van der Meer}
\author{Benjamin Ertel}%
\author{Udo Seifert}%
\affiliation{%
 II. Institut für Theoretische Physik, Universität Stuttgart, 70550 Stuttgart, Germany
}%




\date{\today}

\begin{abstract}
The inference of thermodynamic quantities from the description of an only
partially accessible physical system is a central challenge in stochastic
thermodynamics. A common approach is coarse-graining, which maps the
dynamics of such a system to a reduced effective one. While coarse-graining
states of the system into compound ones is a well studied concept, recent
evidence hints at a complementary description by considering observable
transitions and waiting times. In this work, we consider waiting time 
distributions between two consecutive transitions of a partially observable
Markov network. We formulate an entropy estimator using their ratios to
quantify irreversibility. Depending on the complexity of the underlying network,
we formulate criteria to infer whether the entropy estimator recovers the 
full physical entropy production or whether it just provides a lower bound
that improves on established results. This conceptual approach, which is 
based on the irreversibility of underlying cycles, additionally enables us
to derive estimators for the topology of the network, i.e., the presence
of a hidden cycle, its number of states and its driving affinity. Adopting
an equivalent semi-Markov description, our results can be condensed into a
fluctuation theorem for the corresponding semi-Markov process. This
mathematical perspective provides a unifying framework for the entropy
estimators considered here and established earlier ones. The crucial role of
the correct version of time-reversal helps to clarify a recent debate on 
the meaning of formal versus physical irreversibility. Extensive numerical
calculations based on a direct evaluation of waiting-time distributions
illustrate our exact results and provide an estimate on the quality of the
bounds for affinities of hidden cycles.
\end{abstract}

\maketitle


\section{Introduction}

Over the last two decades, stochastic thermodynamics has emerged as a comprehensive universal framework for describing small driven systems \cite{Sekimoto2010, jarzynski2011, Seifert2012, broeck2015, Peliti2021}. One major paradigm comprises a Markovian, i.e., memoryless dynamics on a set of discrete states, which arises from integrating out fast microscopic degrees of freedom under the assumption of a timescale separation. Such a fairly general Markov network model is of widespread use in the description of chemical and biophysical processes, ranging from chemical reaction networks \cite{Schmiedl2007, Ge2012, Rao2016, ould17, Marsland2019} to protein folding \cite{Bowman2010, Rief2011, Chodera2014}, molecular motors \cite{qian97, Andrieux2006, liepelt2007, chow13, wago16, speck2021} and molecular dynamics in general \cite{Prinz2011, Husic2018, wang2018}.

There is, however, a difference between identifying an effective description of a complex system and actually having full access to it in practice. On the arguably coarsest level of description, one is interested in estimation methods of crucial  quantities like the entropy production. As a prominent result, the thermodynamic uncertainty relation (TUR) \cite{barato2015, ging16, horowitz2020} provides thermodynamic bounds that can be used in estimation techniques for entropy \cite{hwan18, seifert2018, li2019, song2020, vu2020, dech21} or topology \cite{barato2015b, pietzonka2016} if it is possible to measure currents of the underlying system. These currents are a trace of the fundamental time-reversal asymmetry in dissipative systems \cite{kawai2007, parrondo2009} that can also be utilized directly as an entropy estimator \cite{roldan2010, roldan2012, muy2013}. Furthermore, entropy estimators that incorporate or are even based on waiting times between measurable events have been discussed more recently \cite{martinez2019, Skinner2021_1, Skinner2021_2, Ehrich2021}. For a partially visible Markov network,  entropy production can be estimated through the fraction that is visible in the subsystem through passive observation \cite{shiraishi2015} or by controlling adjustable parameters \cite{polettini2017, bisker2017}. 

These methods raise the general issue how an underlying, only partially accessible system is related to a reduced effective model, a topic known as coarse-graining in stochastic thermodynamics. Earlier interest in the field mainly considered  coarse-graining as a mapping in which unresolved  Markov states are lumped into compound states, for example via schemes described in Refs. \cite{Rahav2007, pigolotti2008, Puglisi2010, Knoch2015, seiferth2020}. In general, the resulting system is no longer Markovian, so that a description of the dynamics or the entropy production is formulated in terms of phenomenological, apparent equations \cite{Esposito2012,Mehl2012,Bo2014,Bo2017,seifert2018,Uhl2018}. While particular symmetric systems can be described as semi-Markov processes in this coarse-graining approach \cite{Qian2007, Teza2020, Ertel2021}, a general framework to describe situations with incomplete information remains an open issue. To give a recent example \cite{Hartich2020,Hartich2021}, allowing states that are not contained in any compound state breaks with the well studied paradigm of state lumping as coarse-graining scheme. This novel scheme extends our ability to formulate thermodynamically consistent models while also exhibiting new effects such as kinetic hysteresis that require a refined understanding of the relationship between time-reversal and coarse-graining.

In this work, we discuss thermodynamic inference based on the observation of a few transitions and their waiting time distributions, rather than on the observation of a few states. This strategy has been proposed independently in the very recent Ref. \cite{pedro2022} where the corresponding estimator for entropy production has been introduced and its properties derived using mainly concepts from information theory, in particular, the Kullback-Leibler divergence. In our complementary approach that is based on the analysis of cycles, we will show that the underlying trajectory-dependent quantity obeys a fluctuation theorem. Our analysis reveals that this estimator is the entropy production of a semi-Markov process. In particular, we will show that the description discussed in the present work and in Ref. \cite{pedro2022} shows kinetic hysteresis \cite{Hartich2020}. Mathematically, this effect is the consequence of a time-reversal operation that differs from the one that is usually employed for semi-Markov processes. In this context, higher-order semi-Markov processes \cite{martinez2019} fit into the picture naturally as semi-Markov processes with yet another time-reversal operation. Thus, our mathematical perspective establishes semi-Markov processes as an underlying common model while also highlighting the subtleties involved in identifying the correct time-reversal operation. 

Thermodynamic inference is not limited to estimating entropy production. We show that the waiting time distributions allow us to infer topological properties and further thermodynamic quantities like the number of states in cycles and their driving affinity. Furthermore, we propose an inductive scheme to detect the presence of hidden cycles in a complex network.

The paper is structured as follows. In \autoref{sec:2}, we describe the setup and present our key results qualitatively. The fundamental concepts of our effective description are introduced in \autoref{sec:3} for the paradigmatic model of a single observed link in a unicyclic Markov network. By generalizing these concepts to multicyclic Markov networks in \autoref{sec:4}, we propose and discuss an entropy estimator and inference methods theoretically and numerically. The general framework of multiple observed links in a multicyclic Markov network is discussed in \autoref{sec:5}. In \autoref{sec:6}, we discuss our and related work from the perspective of semi-Markov processes. We conclude with a summary and an outlook on further work in \autoref{sec:7}.

\section{Setup and key qualitative results}
\label{sec:2}

\begin{figure*}[bt]
    \begin{subfigure}[t]{0.22\textwidth}
      \centering
        \includegraphics[width=\linewidth]{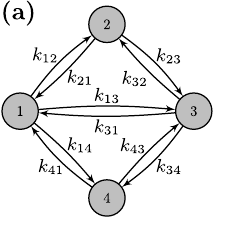}
    \end{subfigure}
    \hfill
    \begin{subfigure}[t]{0.22\textwidth}
      \centering
        \includegraphics[width=\linewidth]{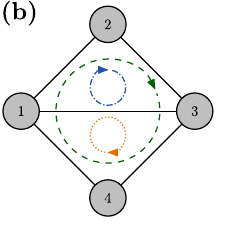}
    \end{subfigure}  
    \hfill
    \begin{subfigure}[t]{0.22\textwidth}
      \centering
        \includegraphics[width=\linewidth]{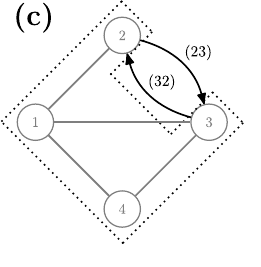}
    \end{subfigure}  
    \hfill
    \begin{subfigure}[t]{0.22\textwidth}
      \centering
        \includegraphics[width=\linewidth]{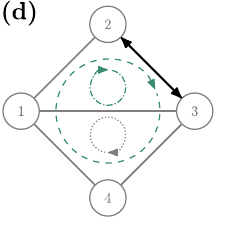}
    \end{subfigure}  
    \begin{subfigure}[t]{0.49\textwidth}
      \centering
        \includegraphics[width=\linewidth]{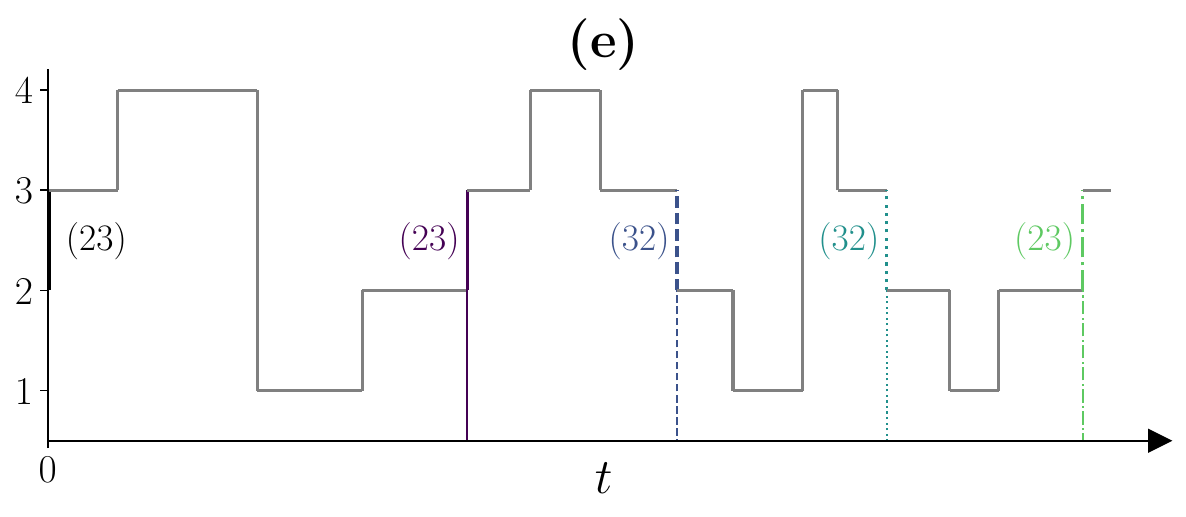}
    \end{subfigure}
    \hfill
    \begin{subfigure}[t]{0.245\textwidth}
        \centering
        \vspace{-3.8cm}
            \includegraphics[width=\linewidth]{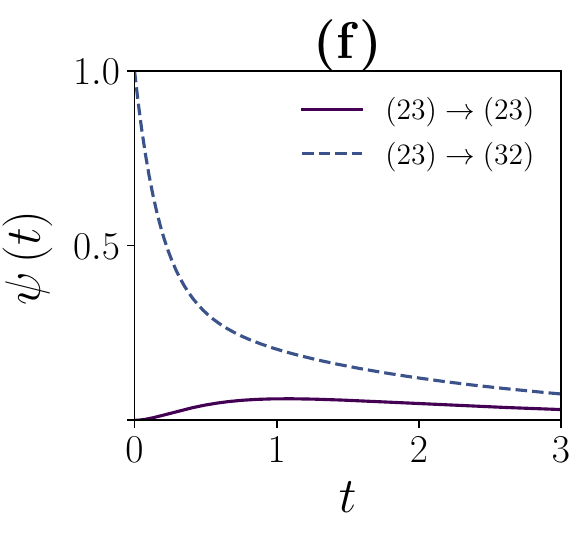}
    \end{subfigure}
    \hfill
    \begin{subfigure}[t]{0.245\textwidth}
        \centering
        \vspace{-3.8cm}
            \includegraphics[width=\linewidth]{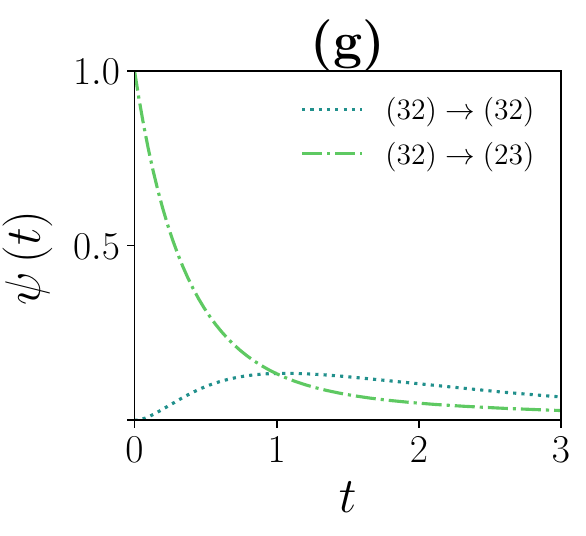}
    \end{subfigure}
    \caption[Example Key]{Key concepts of the effective description for an exemplary Markov network. (a): Markov network including four different states. Every link between state $i$ and state $j$ allows for transitions in both directions with respective transition rates $k_{ij}$ and $k_{ji}$. (b): Different cycles within the network. The three different cycles in the network are numbered incrementally starting with cycle $\mathcal{C}_{0} = (12341)$, drawn as green dashed curve, cycle $\mathcal{C}_{1} = (1231)$, drawn as blue dash-dotted curve and cycle $\mathcal{C}_{2} = (1341)$, drawn as orange dotted curve. By definition in \autoref{Eq:setupAff}, the affinity of $\mathcal{C}_{0}$ is given by $\mathcal{A}_{\mathcal{C}_{0}} = \ln (k_{12}k_{23}k_{34}k_{41}/k_{21}k_{32}k_{43}k_{14})$, $\mathcal{A}_{\mathcal{C}_{1}}$ and $\mathcal{A}_{\mathcal{C}_{2}}$ are defined analogously. Furthermore, these affinities coincide with $\mathcal{A}_{\mathcal{C}} = \ln \mathcal{P(\circlearrowleft)}/\mathcal{P(\circlearrowright)}$, the quotient of probabilities to observe a completed cycle in forward and backward direction, respectively (cf. \autoref{eq:unicyclFluctTheo}). (c): Effective description of the network if only the link between $2$ and $3$ is observable. Observing this link gives information about transitions between $2$ and $3$, i.e. $(23)$ and its reverse $(32)$, and intermediate waiting times. (d): Observable cycles in the effective description. Two successive transitions along the observable link indicate the completion of a cycle. As indicated with gray color, only completions of $\mathcal{C}_{0}$ or $\mathcal{C}_{1}$ can be registered, since $\mathcal{C}_{2}$ does not include the observed link. Additionally, $\mathcal{C}_{0}$ and $\mathcal{C}_{1}$ are drawn as curves with same color because by counting transitions without temporal resolution, we cannot distinguish between both cycles. (e): A trajectory and its effective description. The observable parts of a trajectory of the underlying network are transitions $(23)$ and $(32)$ at corresponding transition times. By conditioning the observed transitions on the previous ones, four different waiting time distributions for the different combinations of subsequent transitions can be defined. (f) and (g): Waiting time distributions for the observable link for fixed transition rates. The four different waiting time distributions of the observed link are illustrated, they were calculated with the method introduced in appendix \ref{App:Absorbing}. The particular choice of transition rates is given in appendix \ref{App:Para}.}
\label{Fig:Key}
\end{figure*}

We start with a general Markov network of $N$ interconnected states, e.g., the one shown in \autoref{Fig:Key}(a). At time $t$, a state $i(t) = k$ is assigned to the physical system, with $k = 1, ..., N$. The time-evolution follows a stochastic description by allowing transitions between two states $k$ and $l$ that are connected by a link, equivalently, an edge, in the network. Quantitatively, these transitions from $k$ to $l$ and their reverse happen instantaneously with transition rates $k_{kl}$ and $k_{lk}$, respectively. We assume that $k_{kl} > 0$ implies $k_{lk} > 0$ to ensure thermodynamic consistency. In the long-time limit $t \to \infty$, the probability $p_k(t)$ to observe the system in a particular state $k$ at time $t$ approaches a constant value $p_k^s$, which characterizes the stationary state of the network.  

In general networks, it is possible to walk along closed loops. These are accessed systematically from the network by identifying its cycles $\mathcal{C}$, which are defined as closed, directed loops without self-crossings. From a thermodynamic perspective, cycles are a crucial concept due to their possibility to break time-reversal symmetry by favoring the forward direction over the reverse or vice versa. This preference is quantified by the cycle affinity $\mathcal{A}_\mathcal{C}$, defined as the product over all forward rates in $\mathcal{C}$ divided by the corresponding backward rates,
\begin{equation}
    \mathcal{A}_\mathcal{C} = \ln \prod _{(kl) \in \mathcal{C}}  \frac{k_{kl}}{k_{lk}}
    \label{Eq:setupAff}
.\end{equation}
As shown in \autoref{Fig:Key}(b), the network from \autoref{Fig:Key}(a) has three different cycles with different affinities. The affinity $\mathcal{A}_\mathcal{C}$ is also related to the entropy production associated with the cycle $\mathcal{C}$ \cite{schnakenberg1976,Hill1989}. For biochemical reactions or driving along a periodic track by a force, the affinity is given by the free energy change or dissipated work, respectively \cite{Seifert2012}.

Cycles $\mathcal{C}$ with non-vanishing affinities give rise to macroscopic, sustained flows along their constituent links, even in the limit of large observation times $T$. These circular flows are the cause of the mean entropy production rate
\begin{equation}
    \braket{\sigma} = \sum_{\mathcal{C}} j_\mathcal{C} \mathcal{A}_\mathcal{C},
    \label{eq:setup:sigma}
\end{equation}
where $j_\mathcal{C}$ is the expected net number of completed cycles\footnote{Typically, the expected net number of completed cycles $j_\mathcal{C}$ does not coincide with any current through an edge contained in the cycle. In particular, $j_\mathcal{C}$ does not rely on the identification of spanning trees or fundamental cycles but is well-defined after specifying $\mathcal{C}$ in a particular network as explained in Ref. \cite{Jiang2004}.} $\mathcal{C}$ divided by the observation time $T$ in the limit $T \to \infty$  \cite{Hill1989, Jiang2004}. If $\braket{\sigma} > 0$, there is a constant rate of dissipation in the stationary state, which is then referred to as a non-equilibrium stationary state (NESS).

Calculating the entropy production via \autoref{eq:setup:sigma} requires the ideal case of knowing all cycles and all cycle currents, which is not practically feasible in general. In our setup, we assume that an external observer measures individual transitions along a limited number of edges connecting neighboring states in the Markov network. Conceptually, this approach coincides with the transition-based effective description proposed in Ref. \cite{pedro2022}. Notationally, we discern transitions from states by utilizing capital letters $I, J, ...$ and write $I = (kl)$ to express that $I$ is a transition from the Markov state $k$ to the Markov state $l$. An example illustrating this effective description for observable transitions $(23)$ and $(32)$ in the Markov network from \autoref{Fig:Key}(a) is shown in \autoref{Fig:Key}(c) and (d). The central objects of interest for this effective description are waiting time distributions of the form
\begin{equation}
    \psi_{I \to J}(t) \equiv P\left(J; T_{J}-T_{I}=t|I\right)
    \label{eq:setup:psiAsProb}
,\end{equation}
which quantify the probability density that the transition $J$ is measured at time $T_{J} = T_{I} + t$ given that the previous transition $I$ was registered at time $T_{I}$. With transitions $I,J$ replacing states $k,l$, waiting time distributions $\psi_{I \to J}(t)$ are the time-resolved analogue of transition rates $k_{kl}$. \autoref{Fig:Key}(e), (f) and (g) illustrate the concept of waiting time distributions for the effective description in \autoref{Fig:Key}(c) and (d). In the following, we will derive several remarkable results centered around these waiting time distributions and their underlying semi-Markov description, which are summarized here on a qualitative level.
\begin{enumerate}
    \item For a unicyclic network, it is sufficient to determine the $\psi_{I \to J}(t)$ from just one edge in order to infer the affinity of the cycle $\mathcal{C}$ and the exact mean entropy production rate $\braket{\sigma}$ from the ratio of these distributions. We recover this result of Ref. \cite{pedro2022} independently, here based on a microscopic fluctuation theorem from the perspective of network cycles. Since the full entropy production is inferred by this estimator, it beats the TUR, which, in general, does not recover the full entropy production even in a unicycle. 
    \item For a multicyclic network, the same information from just one edge yields the affinity of the shortest cycle, its length and the length of the second-shortest cycle this edge is a part of. Second, it yields a lower bound on the largest cycle affinity contributing to the current through this edge. Finally, it provides a lower bound on the overall entropy production of the network that coincides with the bound proposed in Ref. \cite{pedro2022}. This bound is shown to be tighter than the entropy estimator in Ref. \cite{polettini2017} while also omitting any assumptions of physical control over system parameters at the observed edge.
    \item If several edges can be observed, the estimator on total entropy production becomes successively tighter. Based on the ratios of the $\psi_{I \to J}(t)$, we establish operational criteria to infer the presence of hidden cycles and hidden entropy production not accounted for by the estimator. 
    \item From a mathematical perspective, observing transitions results in a semi-Markov process. The cycle-based approach of this work and the information-theoretical approach of Ref. \cite{pedro2022} can be seen as equivalent strategies to establish the entropy production of the corresponding semi-Markov process. From this point of view, we relate the proposed entropy estimator to the semi-Markov entropy estimator proposed and discussed in Refs. \cite{martinez2019, Hartich2021preprint,Bisker2022_Reply} and highlight the crucial role of the different time-reversal operations.
\end{enumerate}

\section{Unicyclic network as paradigm}
\label{sec:3}

For an introductory example, we consider a Markov network with only a single cycle $\mathcal{C}$ in its NESS. In this network, we observe a single edge between neighboring states $k$ and $l$ that is part of the cycle. We assume that forward and backward transitions along this edge can be distinguished and denote forward transitions $(kl)$ as $I_+$ and backward transitions $(lk)$ by $I_-$, respectively. 

On the microscopic level, a waiting time distribution of the form $\psi_{I \to J}(t)$ has contributions only from microscopic trajectories $\gamma^t_{I \to J}$ that start with a transition $I$ and end with another one, $J$, after time $t$ without any other observed transition in between. With a microscopic path weight $\mathcal{P}[\gamma]$ for microscopic trajectories $\gamma$, the waiting time distribution can be expressed as
\begin{equation}
     \psi_{I \to J}(t) = \sum_{\gamma_{I \to J}^t} \mathcal{P}[\gamma_{I \to J}^t|I]
    \label{eq:psiPathWeight}
,\end{equation}
which only sums trajectory snippets of the form $\gamma = \gamma^t_{I \to J}$ with a path weight that is conditioned on the first jump $I$ at time $T_I$. For example, the waiting time distribution $\psi_{I_{+} \to I_{+}}(t)$ originates from a trajectory snippet $\gamma_{I_{+}\to I_{+}}^t$ of length $t$ with the jump sequence $\gamma_{I_+ \to I_+}^t = k \to l \to \cdots \to k \to l$. Likewise, $\psi_{I_{-} \to I_{-}}(t)$ arises from $\gamma_{I_- \to I_-}^t = l \to k \to \cdots \to l \to k$. Although the identification in \autoref{eq:psiPathWeight} is reasonable from a practical point of view, its derivation contains some subtleties that are explained in the full proof of \autoref{eq:psiPathWeight} in appendix \ref{App:PsiIdent}. Since $\gamma_{I_- \to I_-}^t$ is the reverse of $\gamma_{I_+ \to I_+}^t$, the logarithmic ratio of the corresponding waiting time distributions, 
\begin{equation}
    a(t) \equiv a_{I_+ \to I_+}(t) \equiv \ln \frac{\psi_{I_+ \to I_+}(t)}{\psi_{I_- \to I_-}(t)}
    \label{eq:aDef}
,\end{equation}
is a natural, antisymmetric measure of irreversibility of the underlying trajectory. As a first main result, we will show that $a(t)$ is independent of $t$, and, in particular, can be identified with the cycle affinity $\mathcal{A}_\mathcal{C}$,
\begin{equation}
    a_{I_+ \to I_+}(t) \equiv a = - a_{I_- \to I_-}(t) = \mathcal{A}_{\mathcal{C}}
    \label{eq:aIsAffinity}
.\end{equation}
This relation can be seen as a fluctuation theorem applied to sections of the underlying trajectory on the Markov network that give rise to a waiting time distribution $\psi_{I_+ \to I_+}(t)$. These sections are trajectory snippets $\gamma_{I_+ \to I_+}^t$ of the form given above, where the time difference between both jumps $k \to l$ is exactly $t$. To observe the genuine time-reverse $\psi_{I_- \to I_-}(t)$, the underlying trajectory must complete the cycle in the reverse direction, which means
\begin{equation}
    \mathcal{P}[\gamma_{I_- \to I_-}^t|I_-] = \mathcal{P}[\gamma_{I_+ \to I_+}^t|I_+] e^{- \mathcal{A}_{\mathcal{C}}}
    \label{eq:unicyclFluctTheo}
\end{equation}
for the path weights of every possible trajectory snippet $\gamma_{I_\pm \to I_\pm}^t$. Since this argument holds true for all trajectories contributing to the waiting time distribution $\psi_{I_+ \to I_+}(t)$, we can sum the left side of \autoref{eq:unicyclFluctTheo} over all $\gamma_{I_- \to I_-}^t$ and the right side of \autoref{eq:unicyclFluctTheo} over all $\gamma_{I_+ \to I_+}^t$ to conclude
\begin{equation}
    \psi_{I_- \to I_-}(t) = \psi_{I_+ \to I_+}(t) e^{- \mathcal{A}_{\mathcal{C}}}
    \label{eq:unicyclFluctTheo_Psi}
\end{equation}
using \autoref{eq:psiPathWeight}. Inserting \autoref{eq:unicyclFluctTheo_Psi} into \autoref{eq:aDef} proves \autoref{eq:aIsAffinity}.

Since $a(t) = a = \mathcal{A}_{\mathcal{C}}$ is time-independent, we get from \autoref{eq:aDef} to
\begin{equation}
    \mathcal{A}_{\mathcal{C}} = a = \ln \frac{\int _0 ^\infty dt \; \psi_{I_+ \to I_+}(t)}{\int _0 ^\infty dt \; \psi_{I_- \to I_-}(t)} =  \ln \frac{P(I_+|I_+)}{P(I_-|I_-)}
    \label{eq:unicyclQuotient2}
\end{equation}
with an integration over the time $t$. The last equality follows from the definition of $\psi_{I \to J}(t)$ as a joint distribution in $J$ and $t$ in \autoref{eq:setup:psiAsProb}. Thus, the cycle affinity is encoded in conditional probabilities $P(J|I)$ to observe transition $J$ after transition $I$ irrespective of the intermediate waiting time. The relationship between cycle affinities and a time-antisymmetric probability ratio, given by \autoref{eq:aIsAffinity}, or equivalently \autoref{eq:unicyclQuotient2}, indicates that $a(t)$ can be used as an estimator for the mean entropy production rate $\braket{\sigma}$ in the steady state via
\begin{equation}
    \braket{\sigma} = j_\mathcal{C} \mathcal{A}_\mathcal{C} = j_\mathcal{C} a,
    \label{Eq:Sigma_Uni}
\end{equation}
which is exact even for finite observation times $T$, because the average is taken in the NESS. This non-invasive estimator is directly accessible from an operational point of view as by definition $j_\mathcal{C}$ can be calculated by counting transitions along the observed link and $a(t)=a$ can be calculated either directly from histogram data for the waiting time distributions using \autoref{eq:aDef} or from conditional probabilities deduced from observed transitions using \autoref{eq:unicyclQuotient2}. This unicyclic result also recovers one of the main results in Ref. \cite{pedro2022}, here using a technique based on the microscopic cycle fluctuation theorem \autoref{eq:unicyclFluctTheo}. Thus, the result additionally addresses the conceptual issue of relating entropy production, cycles and fluctuation theorems that was raised at the end of Ref. \cite{pedro2022}.

Conceptually, the identification $\mathcal{A} = a(t)$ relies crucially on the observation of transitions rather than states. Two subsequent transitions in the same direction imply a completed cycle with associated entropy production, whereas two visits of the same compound state emerging from state lumping in typical coarse-graining strategies do not. As all transitions except for one are invisible in the present partially accessible system, previous state-based coarse-graining approaches would yield a trivial model containing only a single compound state. Note that alternated observed transitions, observing a forward transition after a backward transition or vice versa, can never imply the completion of an underlying cycle. Therefore, it is not surprising that the estimator of the entropy production of a unicyclic network contains only the statistics of two subsequent transitions in the same direction, as observed in Ref. \cite{pedro2022}.

\begin{figure*}[bt]
    \begin{subfigure}[t]{0.22\textwidth}
      \centering
        \includegraphics[width=0.95\linewidth]{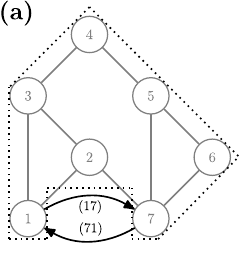}
    \end{subfigure}
    \hfill
    \begin{subfigure}[t]{0.22\textwidth}
      \centering
        \includegraphics[width=0.95\linewidth]{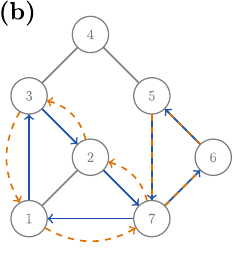}
    \end{subfigure}  
    \hfill
    \begin{subfigure}[t]{0.245\textwidth}
      \centering
        \includegraphics[width=\linewidth]{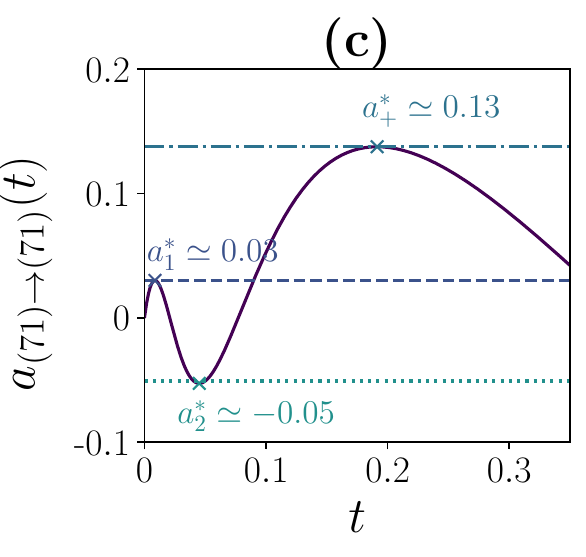}
    \end{subfigure} 
    \begin{subfigure}[t]{0.245\textwidth}
        \centering
            \includegraphics[width=\linewidth]{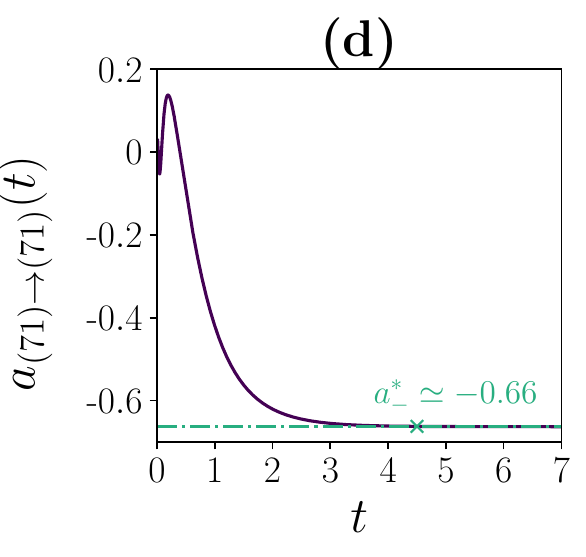}
    \end{subfigure}
    \caption[Multistate Example]{Illustrative example for a partially accessible multicyclic network. (a): Effective description for a seven-state multicyclic network in which the link between state $1$ and state $7$ is observable, leading to $5$ different contributing cycles $\mathcal{C}_{i}$ numbered incrementally. The corresponding transition rates are given in appendix \ref{App:Para}. For cycle $\mathcal{C}_{0} = (1271)$ the affinity $\mathcal{A}_{\mathcal{C}_{0}}$ vanishes, the affinity of cycle $\mathcal{C}_{1} = (13271)$ is $\mathcal{A}_{\mathcal{C}_{1}} = 3.18$, the affinity of cycle $\mathcal{C}_{2} = (134571)$ is $\mathcal{A}_{\mathcal{C}_{2}} = -1.43$, the affinity of cycle $\mathcal{C}_{3} = (1345671)$ is $\mathcal{A}_{\mathcal{C}_{3}} = 7.27$ and the affinity of cycle $\mathcal{C}_{4} = (1234571)$ is given by $\mathcal{A}_{\mathcal{C}_{4}} = -5.61$. (b): Example for a trimmed path. For the snippet $\gamma_{I_+ \to I_+}^t$ depicted with blue arrows, the sequence of visited states is $(713276571)$. The trimmed path for this snippet is $(713271)$  (cf. algorithm in the main text). The corresponding $\gamma_{I_- \to I_-}^t$ is not the reversed sequence but rather $(176572317)$ and depicted with dashed orange arrows. Thus, the associated cycle is $\mathcal{C}_1$, i.e. $\mathcal{P}[\gamma_{I_+ \to I_+}^t|I_+]/\mathcal{P}[\gamma_{I_- \to I_-}^t|I_-] = \mathcal{A}_{\mathcal{C}_1}$. Terms due to the extra loop $(7567)$ cancel in this path weight quotient. (c) and (d): Estimation of the cycle affinities of the contributing cycles based on the extreme values of $a_{(71)\to(71)}(t)$. The maximal value $a^{*}_+ \simeq 0.13$ and the minimal value $a^{*}_- \simeq -0.66$ of $a_{(71)\to(71)}(t)$ are lower and upper bounds for the maximal affinity $\mathcal{A}_{\mathcal{C}_{3}} = 7.27$ and the minimal affinity $\mathcal{A}_{\mathcal{C}_{4}} = -5.61$, respectively. The initial value $a_{(71)\to(71)}(0) = a^{*}_0 = 0$ equals the affinity $\mathcal{A}_{\mathcal{C}_{0}} = 0$ of the shortest network cycle. The local maximum $a^{*}_1 \simeq 0.03$ and the local minimum $a^{*}_2 \simeq -0.05$ can be identified as lower and upper bounds for the affinities $\mathcal{A}_{\mathcal{C}_{1}} = 3.18$ and $\mathcal{A}_{\mathcal{C}_{2}} = -1.43$ of the remaining contributing cycles $\mathcal{C}_{1}$ and $\mathcal{C}_{2}$.}
\label{Fig:Multistate}
\end{figure*}

\section{Multicyclic networks with one observed transition}
\label{sec:4}
For a general network topology, we cannot reconstruct a unique underlying path contributing to the waiting time distributions $\psi_{I_{+} \to I_{+}}(t)$ and $\psi_{I_{-} \to I_{-}}(t)$ as in the unicyclic case. Topologically distinct hidden pathways may result in the same pair of consecutive observed transitions. Nevertheless, bounds for the affinities of those cycles that include the observable link can be derived from the ratio $a(t)$. \add{In addition,} the cycle lengths of specific cycles can be inferred from the short-time limit of the waiting time distributions. Furthermore, the entropy estimator for unicyclic networks can be generalized to the multicyclic case.  

\subsection{Bounds on cycle affinities}
\label{sec:4a}
For each possible underlying cycle $\mathcal{C}$ with $I_+ \in \mathcal{C}$, \autoref{eq:unicyclFluctTheo} is valid with corresponding cycle affinity $\mathcal{A}_\mathcal{C}$, if $\gamma_{I_+ \to I_+}^t$ completes the cycle once in forward direction without taking detours and $\gamma_{I_- \to I_-}^t$ denotes the corresponding reverse path. Thus, the bound
\begin{equation}
    \min _{\mathcal{C}, I_+ \in \mathcal{C}} \mathcal{A}_\mathcal{C} \leq \ln \frac{\mathcal{P}[\gamma_{I_+ \to I_+}^t|I_+]}{\mathcal{P}[\gamma_{I_- \to I_-}^t|I_-]} \leq \max _{\mathcal{C}, I_+ \in \mathcal{C}} \mathcal{A}_\mathcal{C} 
    \label{Eq:Abound} 
\end{equation}
is an immediate consequence for these trajectories $\gamma_{I_+ \to I_+}^t$ by comparing with the smallest and largest possible affinity, respectively. Remarkably, the inequality in \eqref{eq:unicyclFluctTheo} holds true for general $\gamma_{I_+ \to I_+}^t$, if the corresponding $\gamma_{I_- \to I_-}^t$ is defined appropriately by the following algorithm:
\begin{enumerate}
    \item Consider the sequence of states in $\gamma_{I_+ \to I_+}^t$.\\ For $I_+ = (kl)$, this is $(kl\cdots kl)$.
    \item Remove the first and last state:\\ $(kl\cdots kl) \mapsto (l\cdots k)$.
    \item Reading from left to right, remove all closed loops, i.e., as soon as a state $m$ appears twice, remove the intermediate part: $(\cdots amb\cdots cmd\cdots) \mapsto (\cdots amd\cdots)$.
    \item The remaining \emph{trimmed path} visits each state at most once. This trimmed path completed with $I_+$ gives rise to a contributing cycle.
    \item Reverse the trimmed path and reintegrate the first and last state: $(k\cdots l) \mapsto (lk\cdots lk)$.
    \item Reintegrate the closed loops from step $3$ \emph{without} reversing: $(\cdots dma\cdots) \mapsto (\cdots dmb\cdots cma\cdots)$. The resulting sequence of states determines the partial reverse $\mathcal{R}\gamma_{I_+ \to I_+}^t$, which is of the form $\gamma_{I_- \to I_-}^t$.
\end{enumerate}

This procedure identifies a trimmed path of $\gamma_{I_+ \to I_+}^t$ that visits each state at most once. By reversing only this trimmed path, one obtains the partial reverse of $\gamma_{I_+ \to I_+}^t$, which is denoted by $\mathcal{R}\gamma_{I_+ \to I_+}^t$. The associated cycle containing the transition $I_+$ that is reversed by $\mathcal{R}$ has to be one of the possible $\mathcal{C}$ in \autoref{Eq:Abound}. For an example of this procedure, see \autoref{Fig:Multistate}(b). Thus, inverting only the trimmed part of $\gamma_{I_+ \to I_+}^t$ while maintaining the original direction of the remaining transitions restores the inequality in \eqref{eq:unicyclFluctTheo} and hence also the bound in \eqref{Eq:Abound} for every possible microscopic trajectory $\gamma_{I_+ \to I_+}^t$ with the corresponding partner $\gamma_{I_- \to I_-}^t = \mathcal{R}\gamma_{I_+ \to I_+}^t$ defined in this way. 

By averaging over all possible trajectory snippets of length $t$, we can combine \autoref{eq:psiPathWeight} with \autoref{Eq:Abound}, which is now valid for all $\gamma_{I_+ \to I_+}^t$ with corresponding partner $\gamma_{I_- \to I_-}^t$ to conclude
\begin{equation}
    \min _{\mathcal{C}, I_+ \in \mathcal{C}} \mathcal{A}_\mathcal{C} \leq \ln \frac{\psi_{I_+ \to I_+}(t)}{\psi_{I_- \to I_-}(t)} \leq \max _{\mathcal{C}, I_+ \in \mathcal{C}} \mathcal{A}_\mathcal{C} 
    \label{Eq:Abound2} 
\end{equation}
for arbitrary $0 < t < \infty$. For this step, it is important to note that the algorithm provides a bijective mapping $\mathcal{R}$ between trajectories of the form $\gamma_{I_+ \to I_+}^t$ and trajectories of the form $\gamma_{I_- \to I_-}^t$. The inverse mapping is given by applying the same algorithm to $\gamma_{I_- \to I_-}^t$ except for reading right-to-left in step 3 to recover the correct sequence of states for $\gamma_{I_+ \to I_+}^t$.

The quotient in \autoref{Eq:Abound2} can be identified as $a(t)$ via \autoref{eq:aDef}. Thus, the extremal values of $a(t)$ can be identified as bounds on the actual cycle affinity in the form
\begin{align}
    \mathcal{A}_{\mathcal{C}+} \equiv \max _{\mathcal{C}} \mathcal{A}_\mathcal{C} & \geq \sup_{0 \leq t < \infty} a(t)\equiv a^{*}_+\label{Eq:A_m1} \\
    \mathcal{A}_{\mathcal{C}-} \equiv \min _{\mathcal{C}} \mathcal{A}_\mathcal{C} & \leq \inf _{0 \leq t < \infty} a(t)\equiv a^{*}_-
    \label{Eq:A_m}
.\end{align}
Here, maximum and minimum of the affinities are taken over all cycles $\mathcal{C}$ contributing to the observed link. Strong driving along or against the observed link manifests itself in a high positive or negative affinity for a given cycle, respectively. The inequalities \eqref{Eq:A_m1} and \eqref{Eq:A_m} allow us to infer such a source of strong driving from its impact on $a(t)$ from the viewpoint of the observed link. The derived bounds for the cycle affinities are illustrated in \autoref{Fig:Multistate}. \autoref{Fig:Multistate}(c) and (d) show that the extremal affinities $\mathcal{A}_{\mathcal{C}+}$ and $\mathcal{A}_{\mathcal{C}-}$ of the contributing cycles are indeed bounded by the maximum value $a^{*}_+$ and the minimum value $a^{*}_-$ of $a(t)$. Furthermore, the affinity $\mathcal{A}_{\mathcal{C}_{0}}$ of the shortest contributing cycle is always equal to the initial value $a^{*}_0 \equiv a(t=0)$ as we will prove in the following section.

\begin{figure*}[bt]
    \begin{subfigure}[t]{0.24\textwidth}
      \centering
        \vspace{-4.25cm}
        \includegraphics[width=\linewidth]{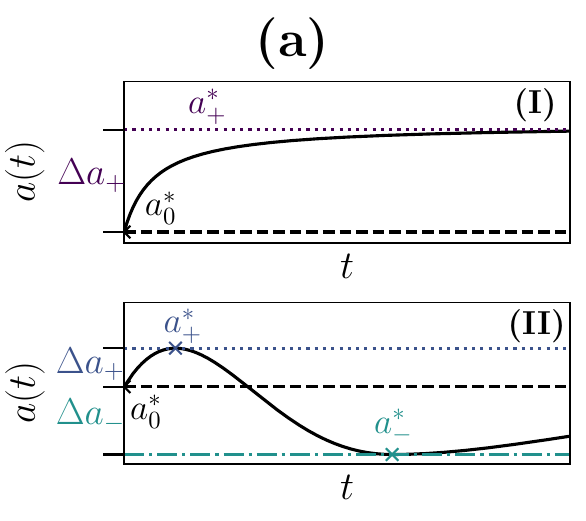}
    \end{subfigure}
    \hfill
    \begin{subfigure}[t]{0.2445\textwidth}
      \centering
        \includegraphics[width=\linewidth]{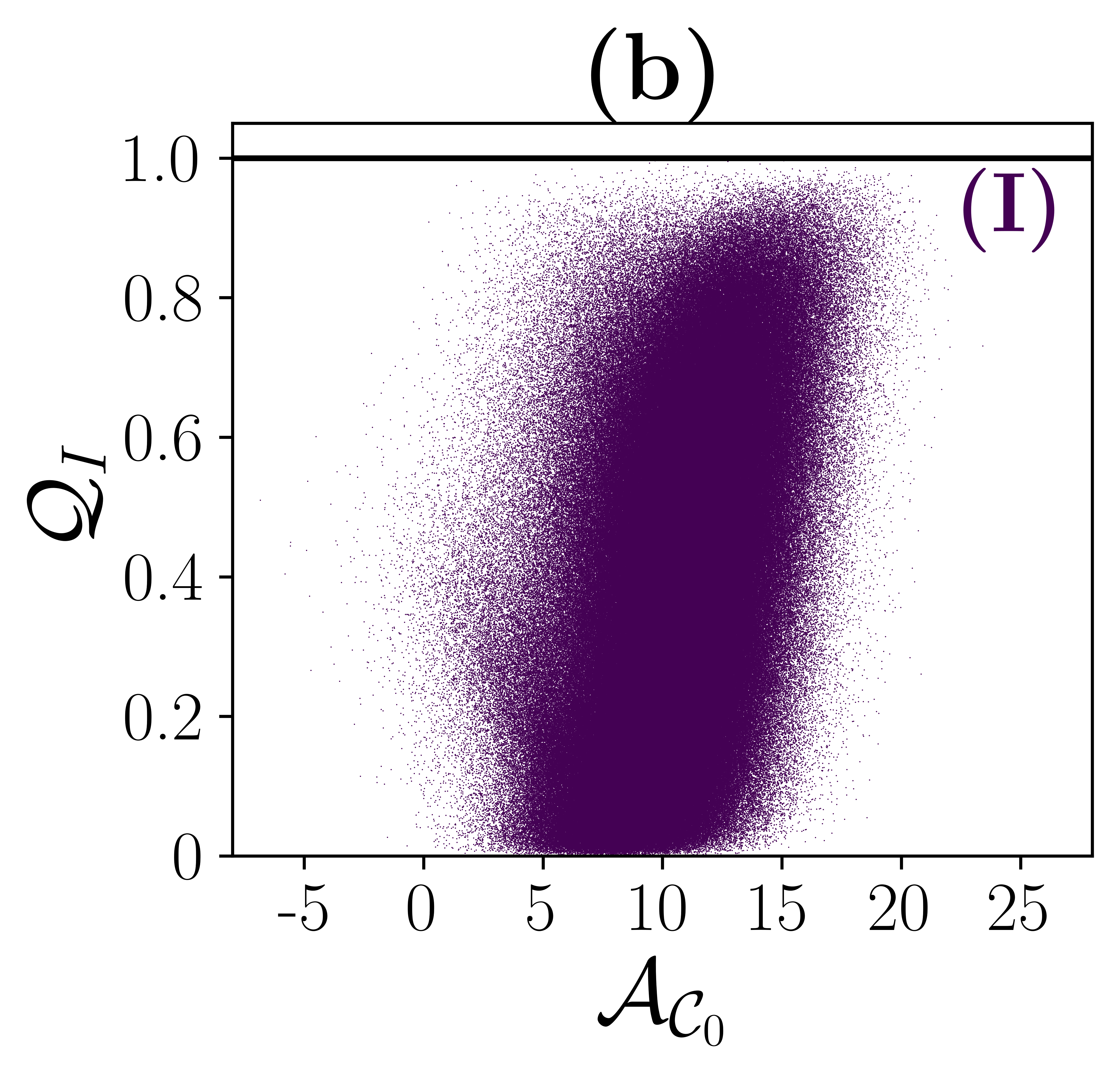}
    \end{subfigure}  
    \hfill
    \begin{subfigure}[t]{0.2495\textwidth}
      \centering
        \includegraphics[width=\linewidth]{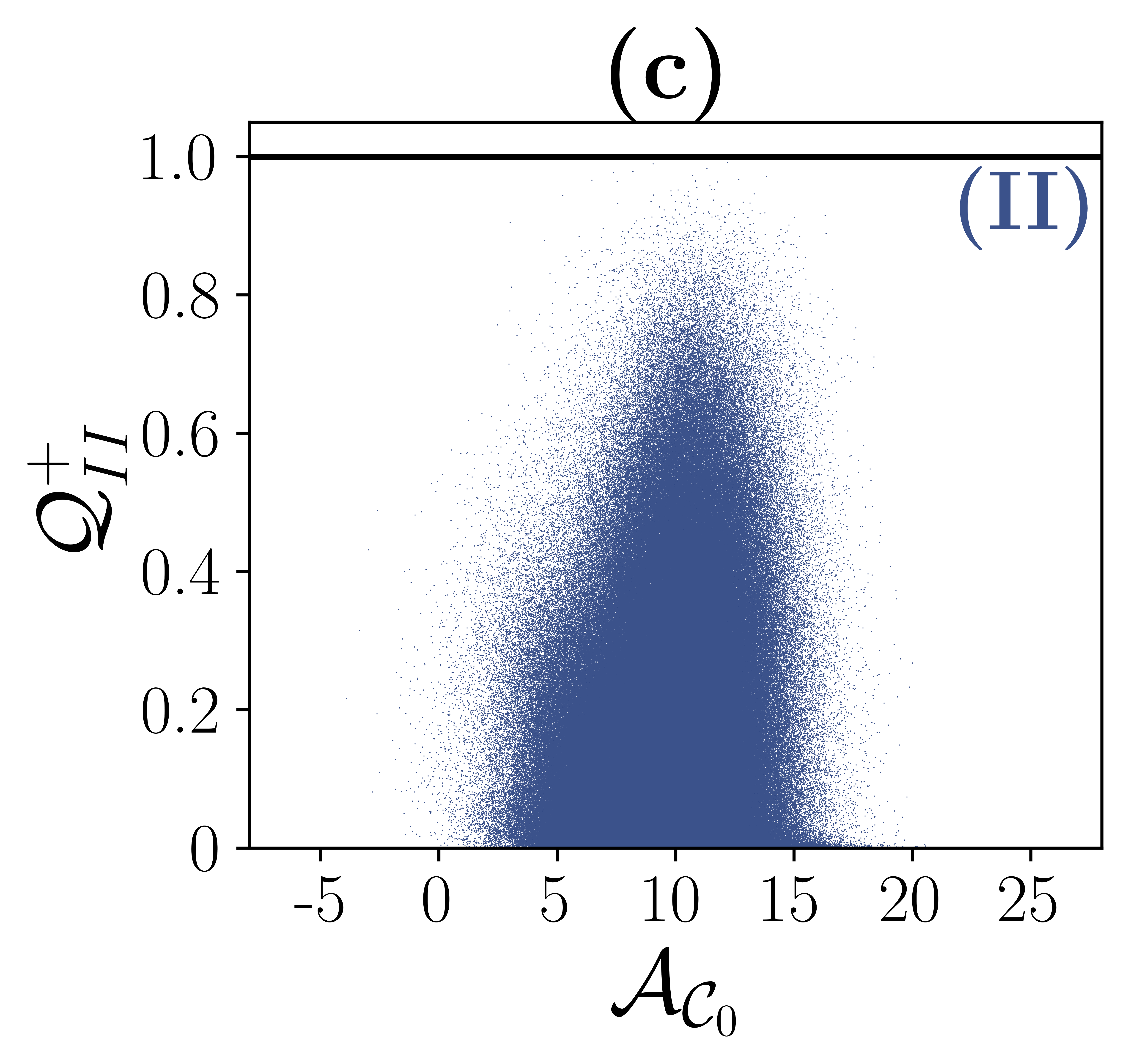}
    \end{subfigure} 
    \begin{subfigure}[t]{0.2495\textwidth}
        \centering
            \includegraphics[width=\linewidth]{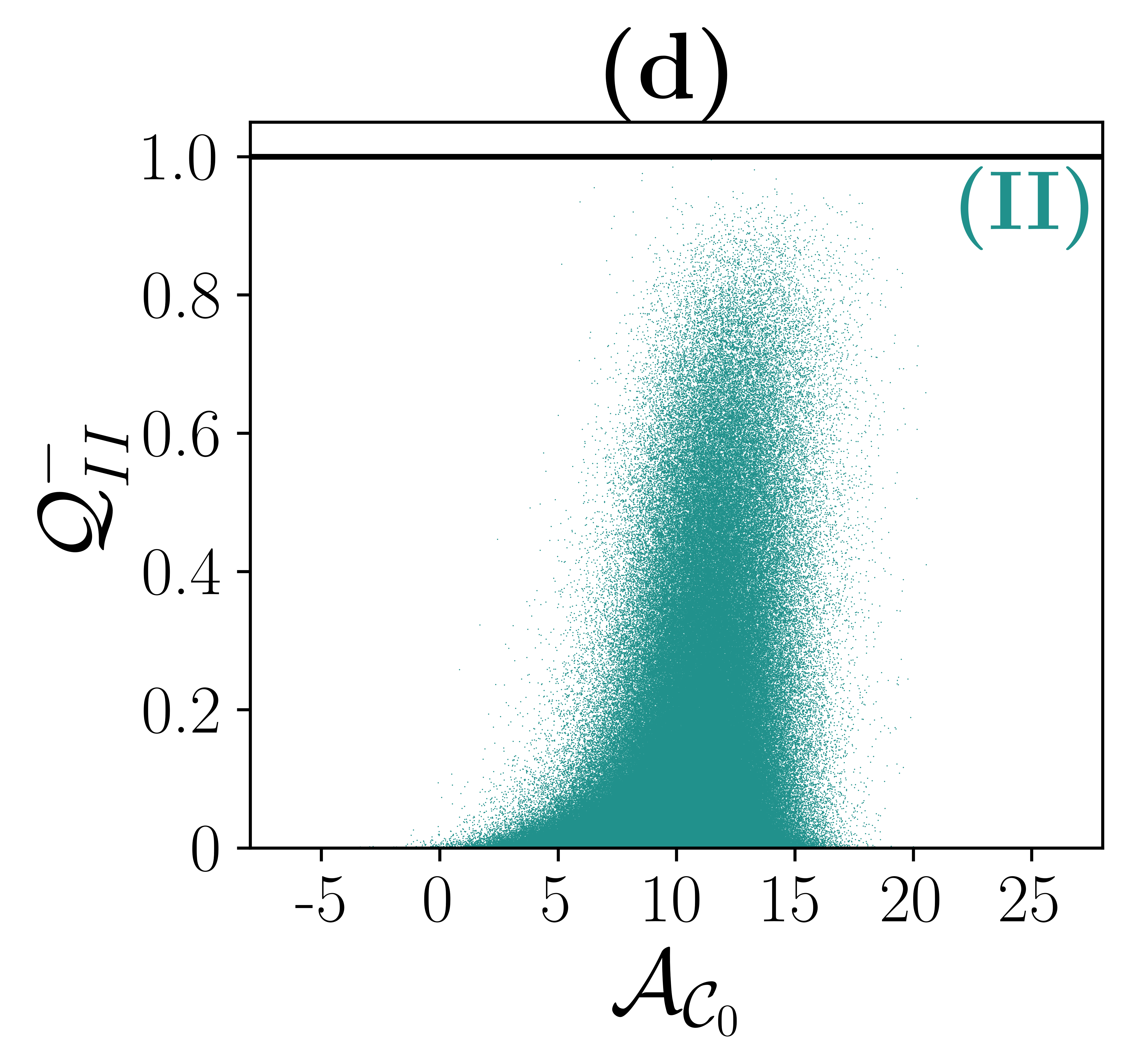}
    \end{subfigure}
    \caption[Scatter]{Quality of the affinity bounds for the seven-state multicyclic network from \autoref{Fig:Multistate}. (a): Illustration of the quantities entering the definition of the quality factors for the two classes of network realizations. (I) shows $a(t)$ for a network realization belonging to class I. Since $a^{*}_0$ is the global minimum of $a(t)$, the quality factor $\mathcal{Q}_I$ for this realization is defined according to \autoref{Eq:Qp_1} with $\Delta a_+ = |a^{*}_+ - a^{*}_0|$ from \autoref{Eq:Delta+}. (II) shows $a(t)$ for a network realization belonging to class II. The quality factors $\mathcal{Q}^{+}_{II}$ and $\mathcal{Q}^{-}_{II}$ for this realization are defined according to \autoref{Eq:Q+} and \autoref{Eq:Q-} with $\Delta a_+ = |a^{*}_+ - a^{*}_0|$ and $\Delta a_- = |a^{*}_- - a^{*}_0|$ from \autoref{Eq:Delta+} and \autoref{Eq:Delta-} respectively. (b), (c) and (d): Quality factors $\mathcal{Q}_I$, $\mathcal{Q}^{+}_{II}$ and $\mathcal{Q}^{-}_{II}$ for 2063495 randomly drawn rate configurations of the multicyclic network as a function of the affinity $\mathcal{A}_{\mathcal{C}_{0}}$ of the smallest contributing cycle. The mean value of the quality factors $\mathcal{Q}_I$ in (b) is given by $\mathcal{Q}_{I}\simeq 0.4$, whereas the mean values of the quality factors $\mathcal{Q}^{+}_{II}$ in (c) and $\mathcal{Q}^{-}_{II}$ in (d) are given by $\mathcal{Q}^{+}_{II}\simeq 0.2$ and $\mathcal{Q}^{-}_{II}\simeq 0.1$ respectively. The difference between the quality factors in (c) and (d) for the same class of network realizations is caused by the ensemble for the transition rates that is biased towards positive affinities, explained in detail in appendix \ref{sec:app:modelparameters}. All quality factors were determined from the corresponding waiting time distributions derived with the method explained in appendix \ref{App:Absorbing}.}
\label{Fig:Scatter}
\end{figure*}

To quantify the quality of the bounds in \eqref{Eq:A_m1} and \eqref{Eq:A_m} for the network from \autoref{Fig:Multistate}(a), we distinguish two different cases of network realizations. A network with a particular configuration of transition rates belongs to class I if the initial value $a^{*}_0$ of $a(t)$ is a global maximum or minimum. An exemplary $a(t)$ of a realization of the network belonging to this class is shown in \autoref{Fig:Scatter}(a), case (I). For this class of network realizations, \autoref{Eq:A_m1} and \autoref{Eq:A_m} only provide a single bound either for the maximal or the minimal affinity of the cycles contributing to the observed link. The other bound is satisfied by the shortest cycle with affinity $a^{*}_0 = \mathcal{A}_{\mathcal{C}_{0}}$. Class II contains the remaining realizations of the network in which $a^{*}_0 = a(t=0)$ is not the global maximum or minimum. An example for an $a(t)$ sorted into class II is given in \autoref{Fig:Scatter}(a), case (II); another one was already shown in \autoref{Fig:Multistate}(c) and (d). For this class of network realizations, equations \autoref{Eq:A_m1} and \autoref{Eq:A_m} provide bounds for both the maximal and the minimal affinity of the cycles contributing to the observed link, respectively.

For both classes of rate configurations, quality factors $\mathcal{Q}$ can be defined such that for $\mathcal{Q}=1$ equality in \autoref{Eq:A_m1} and \autoref{Eq:A_m} holds and the value of the bounds equals the actual affinity of the cycle. For $\mathcal{Q}<1$, the quality factor quantifies the ratio between the value of the bounds and the actual affinity of the corresponding cycle. Using the affinity $\mathcal{A}_{\mathcal{C}_{0}}$ of the shortest cycle given by $a^{*}_0$ as baseline, we introduce the relative distance 
\begin{equation}
  \Delta a(t)\equiv |a(t) - \mathcal{A}_{\mathcal{C}_{0}}| = |a(t) - a^{*}_0|.
  \label{Eq:Delta}   
\end{equation}
The quality factors are defined by comparing the maximal value 
\begin{equation}
  \Delta a_+ = |a^{*}_+ - \mathcal{A}_{\mathcal{C}_{0}}|
  \label{Eq:Delta+}   
\end{equation}
and the minimal value
\begin{equation}
  \Delta a_- = |a^{*}_- - \mathcal{A}_{\mathcal{C}_{0}}|
  \label{Eq:Delta-}   
\end{equation}
of \autoref{Eq:Delta} with the respective actual distance between the true cycle affinities given by $|\mathcal{A}_{\mathcal{C}\pm}-\mathcal{A}_{\mathcal{C}_{0}}|$.

For network realizations belonging to class I, either \autoref{Eq:A_m1} or \autoref{Eq:A_m} is a bound for the affinity of a single cycle. If the initial value $a^{*}_0$ is a global minimum, the maximal affinity $\mathcal{A}_{\mathcal{C}+}$ of the cycles contributing to the observed link is bounded by \autoref{Eq:A_m1}. Thus, the quality factor $\mathcal{Q}_{I}$ for this network realization is defined as
\begin{equation}
    \mathcal{Q}_{I}\equiv \frac{\Delta a_+}{|\mathcal{A}_{\mathcal{C}+}-\mathcal{A}_{\mathcal{C}_{0}}|}.
    \label{Eq:Qp_1}
\end{equation}
If the initial value $a^{*}_0$ is a global maximum, the minimal affinity $\mathcal{A}_{\mathcal{C}-}$ of the cycles contributing to the observed link is bounded by \autoref{Eq:A_m} and the quality factor $\mathcal{Q}_{I}$ for this network realization is given by
\begin{equation}
    \mathcal{Q}_{I}\equiv \frac{\Delta a_-}{|\mathcal{A}_{\mathcal{C}-}-\mathcal{A}_{\mathcal{C}_{0}}|}.
    \label{Eq:Qp_2}
\end{equation}
A graphical illustration of the quantities entering the definition of $\mathcal{Q}_{I}$ is shown in \autoref{Fig:Scatter}(a), case (I).

For network configurations belonging to class II, both \autoref{Eq:A_m1} and \autoref{Eq:A_m} provide nontrivial bounds for the extremal affinities of the contributing cycles. To distinguish both bounds, two quality factors $\mathcal{Q}^{+}_{II}$ and $\mathcal{Q}^{-}_{II}$ defined similarly to \autoref{Eq:Qp_1} and \autoref{Eq:Qp_2} are needed. The quality factor $\mathcal{Q}^{+}_{II}$ defined as
\begin{equation}
    \mathcal{Q}^{+}_{II}\equiv \frac{\Delta a_+}{|\mathcal{A}_{\mathcal{C}+}-\mathcal{A}_{\mathcal{C}_{0}}|},
    \label{Eq:Q+}
\end{equation}
quantities the quality of the bound \autoref{Eq:A_m1} for the maximal affinity $\mathcal{A}_{\mathcal{C}+}$ of the contributing cycles. The quality of the bound \autoref{Eq:A_m} for the minimal affinity $\mathcal{A}_{\mathcal{C}-}$ of the contributing cycles is quantified analogously by
\begin{equation}
    \mathcal{Q}^{-}_{II}\equiv \frac{\Delta a_-}{|\mathcal{A}_{\mathcal{C}-}-\mathcal{A}_{\mathcal{C}_{0}}|}.
    \label{Eq:Q-}
\end{equation}
The quantities entering the definition of $\mathcal{Q}^{+}_{II}$ and $\mathcal{Q}^{-}_{II}$ are illustrated in \autoref{Fig:Scatter}(a), case (II). 

The quality factors for a total of 2063495 randomly drawn realizations of the multicyclic network from \autoref{Fig:Multistate} are shown in \autoref{Fig:Scatter}(b), (c) and (d) as a function of the affinity$\mathcal{A}_{\mathcal{C}_{0}}$ of the smallest contributing cycle. The different structure and mean value of quality factors $\mathcal{Q}_{I}$ for network realizations from class I, shown in \autoref{Fig:Scatter}(b), when contrasted to the structures and mean values of quality factors for network realizations from class II, shown in \autoref{Fig:Scatter}(c) and (d), indicate that the partition into two different classes of network realizations corresponds to distinct features of the network that are reflected in these affinity bounds. The mean value of the quality factors for network realizations belonging to class I is given by $\mathcal{Q}_{I}\simeq 0.4$, which means that the maximal or minimal affinity of the contributing cycles can be estimated based on \autoref{Eq:A_m1} or \autoref{Eq:A_m} with an average accuracy of $0.4$. This result is remarkable because on the one hand, the estimation is based on a non-invasive observation of a single link of the network only and on the other hand, to our knowledge, no coarse-graining inference scheme exists that bounds affinities of a partially accessible network to this degree of precision. The mean values of the quality factors for network realizations belonging to class II are given by $\mathcal{Q}^{+}_{II}\simeq 0.2$ and $\mathcal{Q}^{-}_{II}\simeq 0.1$, respectively. Compared to the bounds for realizations belonging to class I, realizations belonging to class $II$ tend to quantitatively weaker bounds. However, local maxima and minima of $a(t)$ seem to provide further, loose bounds for the affinities of other, non-extremal cycles contributing to the observed link. This numerical finding, illustrated for a given network realization in \autoref{Fig:Multistate}(c), indicates that each successive maximal and minimal value of $a(t)$ corresponds to a contributing cycle. Therefore, the number of successive maximal and minimal values of $a(t)$ can be interpreted as a lower bound for the total number of contributing cycles for networks from class II.

\subsection{Short-time limit and inference of cycle lengths}
\label{sec:4b}

\begin{figure*}[bt]
    \begin{subfigure}[t]{0.22\textwidth}
      \centering
        \includegraphics[width=\linewidth]{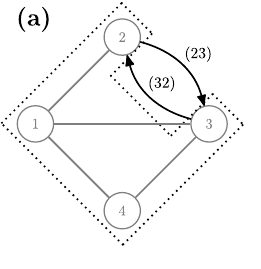}
    \end{subfigure}
    \hfill
    \begin{subfigure}[t]{0.245\textwidth}
      \centering
        \includegraphics[width=\linewidth]{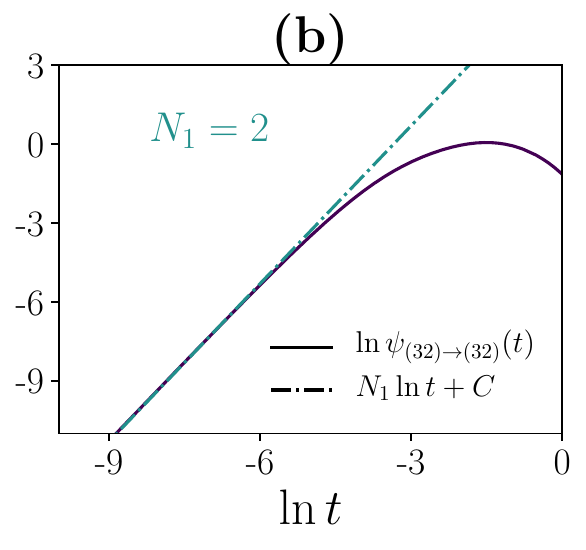}
    \end{subfigure}  
    \hfill
    \begin{subfigure}[t]{0.245\textwidth}
      \centering
        \includegraphics[width=\linewidth]{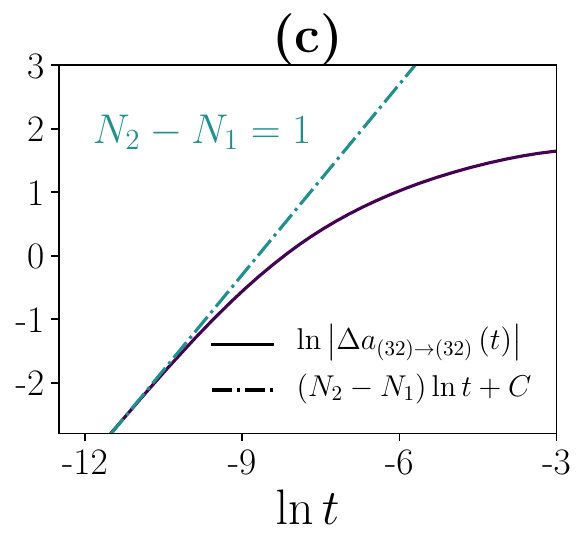}
    \end{subfigure} 
    \begin{subfigure}[t]{0.245\textwidth}
        \centering
            \includegraphics[width=\linewidth]{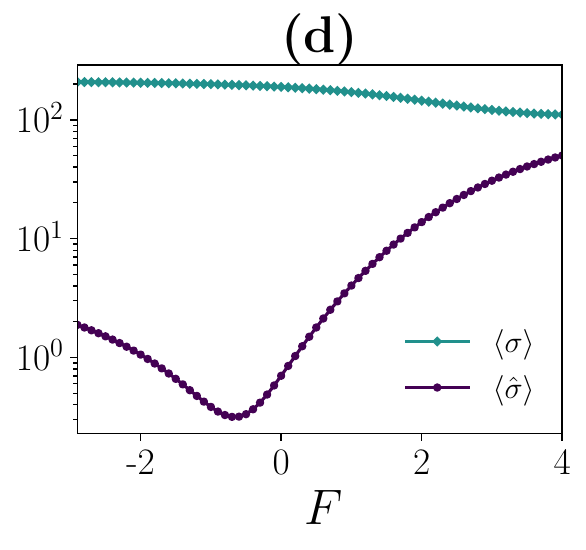}
    \end{subfigure}
    \caption[Diamond Example]{Inference of cycle lengths and entropy estimation for a partially accessible two-cycle network. (a): Effective description of a four-state network with two cycles in which transitions along the link between state 2 and state 3, i.e., $(23)$ and $(32)$, are observable. $F$ is a dimensionless force applied to the observable link between state $2$ and state $3$, all transition rates of the network are given in appendix \ref{App:Para}. (b): Inference of the number of hidden transitions $N_1$ of the smallest network cycle $\mathcal{C}_{0}$ based on waiting time distributions calculated with the method from appendix \ref{App:Absorbing} for fixed $F = \ln 3$. $N_1 = 2$ corresponds to the slope of the short-time limit of $\ln\psi(t)$ resulting in $|\mathcal{C}_{0}| = 3$. (c): Inference of the number of hidden transitions $N_2$ of the second smallest network cycle $\mathcal{C}_{1}$ based on waiting time distributions calculated with the method from appendix \ref{App:Absorbing} for fixed $F = \ln 3$. $N_2 - N_1 = 1$ corresponds to the slope of the short-time limit of $\ln \left|\Delta a(t)\right|$ resulting in $N_{2} = 3$ and $|\mathcal{C}_{1}| = 4$. (d): The estimator $\langle\hat{\sigma}\rangle$ from \autoref{Eq:ccLeqSigma} for the mean entropy production $\langle\sigma\rangle$ of the full network as a function of $F$. The details for the simulations of $\nu_{+|+}(t)$ and $\nu_{-|-}(t)$ are given in appendix \ref{App:Para}. The method from appendix \ref{App:Absorbing} was used to calculate $a(t)$.}
\label{Fig:Diamond}
\end{figure*}

Additional information about the network can be obtained from the time-dependence of the waiting time distributions $\psi_{I_{+} \to I_{+}}(t)$ and $\psi_{I_{-} \to I_{-}}(t)$. In the limit $t\to 0$, only the shortest cycle(s) including the link with forward transition $I_+$ and backward transition $I_-$ contribute(s) to the waiting time distribution, as longer paths lead to effects of higher order in $t$. Thus, we can extract the number of hidden transitions $N_1$ needed to complete the smallest cycle and, if unique, its corresponding affinity $\mathcal{A}_{\mathcal{C}_{0}}$ from the waiting time distributions via
\begin{equation}
    \add{\lim_{t \to 0} \left( t \frac{d}{dt} \ln \psi_{I_{\pm}\to I_ {\pm}}(t) \right) = N_{1}}
    \label{Eq:N1Es}
\end{equation}
and 
\begin{equation}
    \lim_{t\to 0} a_{I_+ \to I_+}(t) = - \lim_{t\to 0} a_{I_- \to I_-}(t) = \mathcal{A}_{\mathcal{C}_{0}}
    \label{Eq:AffS}
,\end{equation}
respectively, as proven in appendix \ref{App:L_Est_N1}. Note that $N_1 + 1$ is equal to the length of the smallest cycle because after $N_1$ hidden transitions, an additional observed transition is needed to complete the full cycle. As an illustration for the identification of $N_1$, we consider the ratio of waiting time distributions for the observable link of the two-cycle network shown in \autoref{Fig:Diamond}(a). \autoref{Fig:Diamond}(b) illustrates that the evaluation of \autoref{Eq:N1Es} for $I_+=(32)$ coincides with $N_1 = 2$, the minimal number of hidden transitions needed to observe (32) after (32) in the smallest cycle of the network. For the multicyclic network of \autoref{Fig:Multistate}, the identification of the affinity in \autoref{Eq:AffS} is illustrated in \autoref{Fig:Multistate}(c) together with the previously discussed affinity bounds, as the affinity $\mathcal{A}_{\mathcal{C}_{0}}$ of the shortest cycle is reflected in the initial value $a_{(71)\to(71)}(0) = 0$. 

Terms of higher order around $t = 0$ of the form $t^N$ encode similar information about cycles with increasing size contributing to the observable link. Qualitatively, we can extract information about the number of hidden transitions $N_2$ needed to complete the second shortest cycle from $a(t)$, since
\begin{equation}
    a(t) - a(0)\sim t^{N_2 - N_1}.
    \label{Eq:aScale}
\end{equation}
More quantitatively and as proven in appendix \ref{App:L_Est_N2}, the absolute value of the relative distance introduced in \autoref{Eq:Delta} can be seen as the lowest order perturbation to the shortest cycle. Typically, e.g., if the affinities of the two shortest cycles do not coincide, this effect is due to the second shortest cycle. In this case, $N_2$ can be extracted from \autoref{Eq:Delta} via 
\begin{equation}
    \add{\lim_{t \to 0} \left( t \frac{d}{dt} \ln \left| \Delta a_{I\to J}(t)\right| \right) = N_{2} - N_{1}}
    \label{Eq:N2}
\end{equation}
if $N_2 > N_1$, i.e., if the shortest cycle is unique. By combining the results from \autoref{Eq:N1Es} and \autoref{Eq:N2}, we can infer $N_{2}$ from observable waiting time distributions. Similar to the length of the shortest network cycle, the length of the second shortest network cycle is given by $N_2 + 1$. \autoref{Fig:Diamond}(c) illustrates the evaluation of \autoref{Eq:N2} for $a(t)$ for $I_{+} = (32)$ leading to $N_{2}-N_{1} = 1$. This result is consistent with $N_{2} = 3$, the number of hidden transitions needed to observe (32) as the next observable transition after (32) along the second smallest cycle of the network.

\subsection{Entropy estimator}
\label{sec:4c}

\subsubsection{Definition}

A time-dependent $a(t)$ implies the presence of a second cycle as longer waiting times between subsequent transitions hint at the completion of longer pathways. Exploiting this time-dependence leads to an entropy estimator that generalizes the estimator of the unicyclic case. To quantify this notion, we let $T$ be the length of a long trajectory with $N+1$ transitions $I_k$ located at $T_{k-1}$. The observation starts with the transition $I_1$ at $T_0 = 0$ and ends with $I_{N+1}$ at time $T_N = T$. Then, the number of subsequent forward or backward transitions with waiting time $t$ in between is given by the time-resolved conditional jump counters defined as
\begin{equation}
    \nu_{+|+} (t) \equiv \frac{1}{T} \sum _{m=1} ^N \delta(T_m - T_{m-1} - t) \delta_{I_{m+1}, I_+} \delta_{I_m, I_+}
    \label{eq:condCount1}
,\end{equation}
with $\nu_{-|-}(t)$ defined accordingly. These time-resolved conditional jump counters are used together with the ratio of waiting times $a(t)$ defined in \autoref{eq:aDef} to define a trajectory-dependent entropy estimator
\begin{equation}
    \hat{\sigma} \equiv \int _0 ^\infty dt\; a(t) \left[ \nu_{+|+} (t) - \nu_{-|-}(t) \right]
    \label{Eq:BEst}
.\end{equation}
Operationally, $\nu_{+|+}(t)$ and $\nu_{-|-}(t)$ can be obtained from counting conditional transitions up to time $t$. $a(t)$ can be obtained from histograms for the waiting time distributions based on waiting times between observed transitions. As proven in appendix \ref{sec:app:entropyestimator} in the limit of long trajectories, i.e., observation times $T \to \infty$, \autoref{Eq:BEst} defines an entropy estimator respecting time-reversal symmetry in thermodynamic equilibrium whose mean additionally satisfies
\begin{equation}
    \braket{\hat{\sigma}} \leq \braket{\sigma}
    \label{Eq:ccLeqSigma}
.\end{equation}
This property can be deduced from a fluctuation theorem
\begin{equation}
    \hat{\sigma} = \lim_{T \to \infty} \frac{1}{T} \ln \frac{\mathcal{P}(\Gamma)}{\mathcal{P}(\widetilde{\Gamma})}
    \label{Eq:fluctTheoCC}
\end{equation}
for the trajectory $\Gamma$ and its time-reverse $\widetilde{\Gamma}$, both emerging from trajectories of the underlying network by a mapping defined by the effective description of the system. An interpretation for $\Gamma$ from a mathematical point of view will be given in \autoref{sec:6}.

\subsubsection{Illustration and comparison to existing methods}
A numerical illustration of the estimator, \autoref{Eq:BEst}, applied to the partially accessible two-cycle network is depicted in \autoref{Fig:Diamond}(d). The mean entropy production $\braket{\sigma}$ and the entropy estimator $\braket{\hat{\sigma}}$ are simulated for long, stationary trajectories and different values of a parameter $F$, which can be interpreted as a driving force applied to the observed link between the states $2$ and $3$. An external observer who is able to tune the force parameter $F$ can find a value for which the net stationary current $0 = j = \int_0 ^\infty dt \; \braket{\nu_{+|+} (t) - \nu_{-|-}(t)}$ vanishes. This set-up and the particular value of $F$ are referred to as stalling conditions and the stalling force, respectively \cite{polettini2017, bisker2017, martinez2019}. Knowing this stalling force through either measurement or calculation amounts to knowing the effective ''pressure`` the remaining network exerts on the link $(23)$ against the force $F$. This information is incorporated in the so-called ''informed partial`` entropy estimator $\braket{\sigma_{IP}}$ introduced in \cite{polettini2017}. Since the remaining network is taken into account through the effective pressure, $\braket{\sigma_{IP}}$ surpasses the estimator obtained by merely measuring the ''passive partial`` entropy production $\braket{\sigma_{PP}}$ that can be attributed to the transitions in an observed subset \cite{shiraishi2015}, i.e. 
\begin{equation}
    \braket{\sigma_{PP}} \leq \braket{\sigma_{IP}} \leq \braket{\sigma}
\end{equation}
as proven in the context of the ''informed partial`` estimator in \cite{bisker2017}.

Under stalling conditions, both estimators $\braket{\sigma_{PP}}$ and $\braket{\sigma_{IP}}$ become trivial, because they cannot rule out the possibility that the underlying system is at equilibrium if $j = 0$. The introduced time-resolved estimator $\braket{\hat{\sigma}}$, however, is able to infer non-equilibrium since $\braket{\hat{\sigma}} > 0$ even if $j = 0$, as additional information enters its definition in \autoref{Eq:BEst}. Intuitively, the waiting time distributions encode information about the hidden cycle in their time-dependence through a non-constant $a(t)$. More quantitatively, the estimator $\braket{\hat{\sigma}}$ defined by \autoref{Eq:BEst} numerically reproduces the bound of the waiting time distribution based estimator proposed in \cite{martinez2019} for the network in \autoref{Fig:Diamond}. Both the estimator in \cite{martinez2019} and $\braket{\hat{\sigma}}$ share the features of considering successive transitions and adding a time-resolution through waiting time distributions. However, $\braket{\hat{\sigma}}$ is formulated without the framework of a higher-order semi-Markov process or a Markov chain decimation scheme. While these differences render a general quantitative comparison with our estimator difficult, $\braket{\hat{\sigma}}$ beats \add{the informed partial estimator} $\braket{\sigma_{IP}}$ for long, stationary trajectories, 
\begin{equation}
    \braket{\sigma_{IP}} \leq \braket{\hat{\sigma}} \leq \braket{\sigma}
    \label{eq:betterThanIP}
,\end{equation}
as we will prove in appendix \ref{Sec:supp:Prove_IP}. Note that the expectation values are still taken in the limit of large observation times in which finite-time effects at the initial and final transition can be neglected. It is also evident from the proof that the equality is achieved in the first relation if and only if $a(t)$ is time-independent. Equality in the second relation is achieved if and only if removing the observed edge results in a network in which detailed balance is satisfied. To give a less formal interpretation of \autoref{eq:betterThanIP}, observational access to the waiting time distributions contains more information than operational access to the observed links via the stalling force $F$. In particular, it is possible to measure $F$ via
\begin{equation}
    - F = \ln \frac{P(I_+|I_+)}{P(I_-|I_-)} = \ln \frac{\braket{\int_0 ^\infty dt \, \nu_{+|+}(t)}/\braket{n_+}}{\braket{\int_0 ^\infty dt \, \nu_{-|-}(t)}/\braket{n_-}},
    \label{Eq:F_IP}
\end{equation}
without perturbing the system at all, as we will prove in appendix \ref{Sec:supp:Prove_IP}. 

\section{Multiple observed links in a multicyclic network}
\label{sec:5}

Access to additional observable transitions provides further information about the underlying network, which allows us to infer topology qualitatively by identifying allowed and forbidden sequences of transitions and quantitatively by sharpening our entropy estimator for multicyclic networks.

\subsection{Entropy estimator}
For $M$ observed links there are $2M$ possible transitions and a $2M\times2M$ matrix of quotients
\begin{equation}
    a_{IJ}(t) \equiv \ln \frac{\psi_{I \to J}(t)}{\psi_{\widetilde{J} \to \widetilde{I}}(t)}
    \label{eq:multicyclQuotient}
\end{equation}
with $I,J \in \{ I_+^{(1)}, I_-^{(1)}, ..., I_-^{(M)} \}$. Here, $\widetilde{I}$ is defined as the reverse transition $\widetilde{I}^{(m)}_\pm \equiv I^{(m)}_\mp$, which yields a skew-symmetry $a_{IJ} = - a_{\widetilde{J}\widetilde{I}}$. Intuitively, the ratio in \autoref{eq:multicyclQuotient} encodes the entropy production term of an effective two-step trajectory $\Gamma_{IJ}^t = I \to J$ of length $t$. This term is related to the path weights of microscopic trajectory snippets $\gamma_{I \to J}^t = k \to l \to \cdots \to o \to p$ of the same length $t$ between two observed transitions $I = (kl)$ and $J = (op)$ in the form
\begin{equation}
    a_{IJ}(t) = \ln \frac{\mathcal{P}[\Gamma_{IJ}^t|I]}{\mathcal{P}[\Gamma_{\tilde{J}\tilde{I}}^t|\widetilde{J}]} = \ln \frac{\sum_{\gamma_{I \to J}^t} \mathcal{P}[\gamma_{I \to J}^t|I]}{\sum_{\gamma_{\tilde{J} \to \tilde{I}}^t} \mathcal{P}[\gamma_{\tilde{J} \to \tilde{I}}^t|\widetilde{J}]}
    \label{eq:multicyclQuotient2}
.\end{equation}
Similar to the unicyclic case in \autoref{eq:aDef}, unobserved degrees of freedom in the microscopic path $\gamma_{I \to J}^t$ are integrated out by the summation over the path weights. The ratios in \autoref{eq:multicyclQuotient} allow us to generalize $\hat{\sigma}$, defined in \autoref{Eq:BEst}, to multiple observed transitions. We define the conditional counters as 
\begin{equation}
    \nu_{J|I} (t) \equiv \frac{1}{T} \sum _{m=1} ^N \delta(T_m - T_{m-1} - t) \delta_{I_{m+1}, J} \delta_{I_m, I}
    \label{eq:condCount2}
,\end{equation}
where we adopt the same notation as in \autoref{eq:condCount1}, i.e., the $m$-th transition $I_m$ is located at $T_{m-1}$. The sum over all $a_{IJ}(t)$ in a trajectory constitutes the entropy estimator
\begin{equation}
    \hat{\sigma} \equiv \sum _{IJ} \int_0^\infty dt\; a_{IJ}(t) \nu_{J|I}(t)
    \label{Eq:CEst}
,\end{equation}
which reduces to \autoref{Eq:BEst} in the case of a single link, i.e., two possible transitions $I_\pm = \pm$. Thus, registering a jump $J$ after a previous jump $I$ during an observation of a long trajectory increases $\hat{\sigma}$ by $a_{IJ}(t)$, an antisymmetric increment in which inaccessible data beyond the registered observable one is integrated out. The entropy estimator is thermodynamically consistent in the sense of \autoref{Eq:ccLeqSigma} and satisfies the fluctuation theorem from \autoref{Eq:fluctTheoCC} in the long-time limit, $T \to \infty$. Moreover, the definition \eqref{Eq:CEst} provides the fluctuating counterpart of the entropy estimator for multicyclic networks introduced in Ref. \cite{pedro2022}, which is given by $\braket{\hat{\sigma}}$ in our notation.

\subsection{Network topology} 
\label{sec:5b}

When we consider multiple transitions, their relative position in the network has crucial impact on the observed data. For a given network, the waiting time distribution $\psi_{I \to J}(t)$ does not only depend on the pair of transitions $I,J$, but the entire set of observed links. For example, in the effective description of the network in \autoref{Fig:Diamond}(a), $a_{(23)(23)}(t)$ is time-dependent but becomes time-independent if in addition the transitions $(13)$ and $(31)$ are observed. The reason is that the fluctuation-theorem-like argument for the affinity can be restored, since observing $\psi_{(23) \to (23)}(t)$ necessarily implies completion of the cycle $\mathcal{C} = (23412)$. Formulated differently, we can retrace the arguments underlying \autoref{Eq:Abound} to deduce an equality
\begin{equation}
    \mathcal{A}_\mathcal{C} = \ln \frac{\mathcal{P}[\gamma_{(23) \to (23)}^t|(23)]}{\mathcal{P}[\gamma_{(32) \to (32)}^t|(32)]}
,\end{equation}
because the only possible completed cycle is $\mathcal{C}$. Based on this observation, we can conclude in more general terms that increasing the number of observed links in a network decreases the possible pathways in the remaining, hidden part of the underlying Markov network. This subnetwork, which is obtained by removing all observed links from the Markov network, will be denoted as \textit{hidden subnetwork}. While the hidden subnetwork is made up of the same states as the Markov network, it contains fewer links and therefore may be disconnected. 

We can make a few technical but far-reaching observations, which are here formulated for long, stationary trajectories, i.e., expectation values are taken in the NESS and in the limit $T \to \infty$, as before. Let $I = (kl)$ and $J = (op)$ be two arbitrary observed transitions in the network.
\begin{enumerate}
    \item If the hidden subnetwork is topologically trivial, i.e. does not contain any cycles, then $\braket{\hat{\sigma}} = \braket{\sigma}$. Moreover, all $a_{IJ}(t)$ are time-independent.
    \item A time-dependent $a_{IJ}(t)$ implies the presence of a cycle in the hidden subnetwork. More precisely, if $a_{IJ}(t)$ is non-constant in time, then there is a cycle with non-vanishing affinity in the hidden subnetwork that connects the Markov states $l$ and $o$. In particular, $\braket{\hat{\sigma}} < \braket{\sigma}$.
    \item If $J$ cannot be an immediate successor of $I$, i.e. if $\psi_{I \to J}(t) = 0$, the Markov states $l$ and $o$ are not connected in the hidden subnetwork. In particular, we can leave out at least one observed transition without decreasing $\braket{\hat{\sigma}}$.
    \item The converse of 2 is not true. It is possible that  $a_{IJ}(t)$ is constant in time despite a cycle with nontrivial affinity containing both $l$ and $o$. However, this behavior is not the generic case but rather requires high symmetry. An explicit example containing such an invisible cycle is provided in appendix \ref{sec:counterexample}.
\end{enumerate}

These four results are based on the microscopic origin of $a_{IJ}(t)$ as a ratio of path weights as indicated in \autoref{eq:multicyclQuotient2}. The crucial argument is an extension of the reasoning used in the unicyclic case to relate ratios of path weights to the cycle affinity $\mathcal{A}$ (cf. \autoref{eq:unicyclFluctTheo}). We consider two consecutive transitions $I = (kl)$, $J = (op)$ and two arbitrary paths $\gamma_1$ and $\gamma_2$ starting and ending in the Markov states $l$ and $o$, respectively. Their path weights satisfy
\begin{equation}
    \ln \frac{\mathcal{P}[\gamma_1|l]}{\mathcal{P}[\tilde{\gamma}_1|o]} - \ln \frac{\mathcal{P}[\gamma_2|l]}{\mathcal{P}[\tilde{\gamma}_2|o]} = \mathcal{A}_{12}
    \label{eq:multiTwoPaths}
,\end{equation}
where $\mathcal{A}_{12}$ is the affinity of the closed loop obtained by appending $\tilde{\gamma}_2$ to $\gamma_1$. If the hidden subnetwork does not contain any cycles, $\mathcal{A}_{12} = 0$ follows trivially. Since $\gamma_1$ and $\gamma_2$ are arbitrary, \autoref{eq:multiTwoPaths} implies the existence of a specific number $a_{IJ}$ satisfying 
\begin{equation}
    \mathcal{P}[\gamma_{I \to J}^t|I] e^{a_{IJ}} = \mathcal{P}[\gamma_{\widetilde{J} \to \widetilde{I}}^t|\widetilde{J}]
    \label{eq:multiFluctTheo}
\end{equation}
for paths $\gamma_{I \to J}^t$ of arbitrary length $t$ with time-reverse $\gamma_{\widetilde{J} \to \widetilde{I}}^t$. By summing the previous equation over all possible trajectories of the form $\gamma_{I \to J}^t$, we conclude
\begin{equation}
    a_{IJ}(t) = \ln \frac{\sum_{\gamma_{I \to J}^t} \mathcal{P}[\gamma_{I \to J}^t|I]}{\sum_{\gamma_{\widetilde{J} \to \widetilde{I}}^t} \mathcal{P}[\gamma_{\widetilde{J} \to \widetilde{I}}^t|\widetilde{J}]} = \ln \frac{\psi_{I \to J}(t)}{\psi_{\widetilde{J} \to \widetilde{I}}(t)}
    \label{eq:multiA}
.\end{equation}
In particular, $a_{IJ}(t)$ is time-independent if the hidden subnetwork does not contain any cycles or if it satisfies detailed balance, i.e., any cycles in the hidden subnetwork have vanishing affinity. This argument establishes rule 1. To emphasize the relation to our previous results, we note that \autoref{eq:multiA} can be seen as a special case of the affinity bounds from \eqref{Eq:Abound}, which collapse to equalities if the set of possible $\mathcal{A}_\mathcal{C}$ contains only one element. If the hidden subnetwork is a spanning tree, the diagonal element $a_{II} = \mathcal{A}_C$ is the affinity of the cycle $\mathcal{C}$ in the unicyclic network obtained by adding the link $I$ back to the hidden subnetwork. In particular, every cycle passes through at least one observed link and has therefore been registered. Since NESS entropy production stems from cycle currents, it seems plausible to conjecture $\braket{\hat{\sigma}} = \braket{\sigma}$. Up to contributions from the first and last transition of the trajectory, the statement even holds on the level of individual trajectories in the form
\begin{equation}
    \hat{\sigma} = \sigma
,\end{equation}
as will be proven in appendix \ref{sec:app:multiProof}.

Rule $2$ is obtained from \autoref{eq:multiTwoPaths} by reversing the argument above. Since a nontrivial time-dependence $a_{IJ}(t)$ is impossible if $\mathcal{A}_{12}$ vanishes for all $\gamma_1$, $\gamma_2$, there must be at least one cycle with non-vanishing affinity. We will now argue that, despite the counterexample given in appendix \ref{sec:counterexample}, the converse of rule $2$ is usually satisfied in a generic set-up. If $a_{IJ}(t)$ is constant in time, it equals its limit $a_{IJ}(0)$ as $t \to 0$. By a timescale separation argument similar to \autoref{Eq:AffS} only the shortest connection between the corresponding Markov states $l$ and $o$ contributes in the short-time limit, whereas longer connections are suppressed and lead to higher-order effects. A hidden cycle containing $l$ and $o$ can be split along these states, giving rise to two topologically distinct pathways $\gamma_1$ and $\gamma_2$. Unless both pathways contain the exact same number of states, one class of paths is suppressed by the other in the short-time limit. Thus, the hidden cycle must contain an even number of states to avoid this timescale separation argument. In addition to this purely qualitative argument, generic choices of transition rates generally lead to different first-passage times from $l$ to $o$ depending on the topology of the path, which would also lead to a non-trivial time dependence in $a_{IJ}(t)$.

While the derivation of rule $3$ is straightforward from a mathematical point of view, it is of high value operationally as it can be used to infer the connected components of the hidden subnetwork. In addition, this rule describes a scheme to identify the transitions needed to recover the full entropy production. While rule 2 gives a simple criterion when a particular set of observed transitions is insufficient to conclude $\braket{\sigma} = \braket{\hat{\sigma}}$, rule 3 formulates a complementary criterion about transitions which are redundant for the entropy estimate. On the level of the Markov network, restoring the minimal number $n$ of observed links $I_1, ..., I_n$ to connect $l$ and $o$ does not create any cycles in the hidden subnetwork. Since entropy production in the steady state is always due to cycle currents, the entropy production in the hidden subnetwork is not increased by not observing $I_1, ..., I_n$, i.e. by adding $I_1, ..., I_n$ to the hidden subnetwork.

The interplay of statement 2 working ''bottom-up`` and statement 3 coming ''top-down`` is not limited to assessing the quality of the discussed estimator $\hat{\sigma}$. It is also an algorithm for inferring topological aspects of the Markov network by identifying underlying spanning trees, connected components, the position of hidden cycles and, lastly, their affinities and lengths by combining these rules with the methods introduced in \autoref{sec:4}.

\section{Unifying semi-Markov perspective}
\label{sec:6}

\subsection{Identification of the semi-Markov description}

In the transition-based description, each trajectory $\zeta$ of the underlying Markov network is mapped to a trajectory $\Gamma$ that includes only the observable transitions and the waiting times in between, i.e., symbolically
\begin{equation}
    \zeta \mapsto \Gamma[\zeta].
    \label{Eq:Map}
\end{equation}
Clearly, this mapping from $\zeta$ to $\Gamma$ is well-defined and many-to-one. Adopting a different yet equivalent perspective, this kind of mapping for the underlying trajectory can be seen as a type of milestoning using the space of observable transitions for partitioning. Milestoning is a particular coarse-graining scheme from molecuar dynamics simulations \cite{Elber2020} introduced to stochastic thermodynamics in \cite{Hartich2020,Hartich2021}. In short, the milestones represent certain events, whose occurrence indicates the crossing of a milestone that updates the coarse-grained state of the system. 

In practice, this approach results in a semi-Markov description for the coarse-grained system defined on the space of observable transitions. In other words, each observed transition $I$ is identified as a state in the semi-Markov model. The following discussion includes the key concepts of semi-Markov processes in the context of stochastic thermodynamics, see \cite{Qian2007,maes2009,Ertel2021} for details. The equivalence of the transition-based description to a semi-Markov model becomes evident on the level of single trajectories emerging from the mapping in \autoref{Eq:Map}. An effective trajectory $\Gamma$ containing $N+1$ transitions starting and ending with registered transitions $I_{1}$ at time $T_0 = 0$ and $I_{N+1}$ at time $T_N = T$, respectively, is fully characterized by the sequence
\begin{equation}
    \Gamma = \{(I_{1},T_1),(I_{2},T_2),...,(I_{N},T_N)\}
    \label{Eq:SmSeq}
\end{equation}
for $0 \leq t < T_N$. From a mathematical point of view, the sequence in \autoref{Eq:SmSeq} precisely defines a particular realization of a semi-Markov trajectory \cite{Qian2007}, in which the $\{ I_k \}$ take the role of the states. Compared to a Markov process, in which the system is fully described by specifying the state $i$, a full semi-Markov description of the system requires knowing the state $I$ and the waiting time $t$ that has elapsed since $I$ has been entered.

\subsection{Semi-Markov kernels and embedded Markov chain}

Since the theory of semi-Markov processes provides the mathematical framework of the effective description, quantities defined for the latter can be expressed in the language of corresponding semi-Markov processes. The waiting time distribution $\psi_{I\to J}(t)$ assigned to each transition $I$, dubbed as inter-transition time density in Ref. \cite{pedro2022}, is called the semi-Markov kernel in this framework. A semi-Markov kernel $\psi_{I\to J}(t)$ is defined as the joint distribution of waiting time $t$ and transition destination $J$ if the actual state is $I$ with age zero, which coincides precisely with the definition of the waiting time distributions in \autoref{eq:setup:psiAsProb}. Integrating out the waiting time $t$ of a semi-Markov kernel results in conditional probabilities
\begin{equation}
    p_{IJ} \equiv P(J|I) = \int_0^{\infty} dt\, \psi_{I\to J}(t)
    \label{Eq:SemiPTrans}
\end{equation}
for a transition between two semi-Markov states irrespective of the waiting time in $I$. These probabilities, whose ratios have already been used in \autoref{eq:unicyclQuotient2}, can now be placed in a mathematical context. Based on the transition probabilities $p_{IJ}$ defined by \autoref{Eq:SemiPTrans}, the concept of the embedded Markov chain (EMC) can be established for every semi-Markov process by integrating out its time variable \cite{Qian2007}. The embedded Markov chain of the effective trajectory in \autoref{Eq:SmSeq} is given by the sequence
\begin{equation}
    \Gamma_{\text{EMC}} = (I_{1},I_{2},...,I_{N+1})
    \label{Eq:SmEMC}
\end{equation}
of observed transitions. The transition probabilities of the corresponding discrete-time Markov process are given by \autoref{Eq:SemiPTrans}. 

\subsection{Path weight and time-reversal operation}

According to the semi-Markov description, the path weight $\mathcal{P}[\Gamma|I_1,0]$ of the effective trajectory $\Gamma(t)$ conditioned on the first transition is simply given by
\begin{equation}
    \mathcal{P}[\Gamma|I_1,0] = \prod_{i=1}^{N}\psi_{I_{i}\to I_{i+1}}(t_i),
    \label{Eq:SMPW}
\end{equation}
with $t_i = T_i - T_{i-1}$, where we follow the conventional definition \cite{chari1994,Girardin2003,Qian2007,maes2009}. \autoref{Eq:SMPW} coincides with the effective path weight defined for trajectories of the transition-based description in Ref. \cite{pedro2022}. Note that the first and last transition do not need to be treated differently \cite{Qian2007,maes2009,Esposito2008,Ertel2021}, since the trajectory starts and ends with a transition by construction.

The time-reversal operation for the present semi-Markov process is not given by the conventional time-reversal operation for semi-Markov processes. Instead of simply reversing $\Gamma$ in time, as proposed in \cite{chari1994} and \cite{Qian2007}, two peculiarities emerging from the time-reversal of the underlying trajectory $\zeta$ have to be taken into account. First, $\Gamma$ contains observed transitions that are odd under time-reversal similar to momenta and therefore need to be reversed \cite{martinez2019,Hartich2020,Crutchfield2022}. Thus, it is natural to define the reversed transition $\widetilde{I}$ for a transition $I$ as
\begin{equation}
    I = (kl)\;\rightarrow\;\widetilde{I} \equiv (lk).
\end{equation}
Second, we observe an effect introduced as \emph{kinetic hysteresis} in \cite{Hartich2020}. After registering a transition $I = (ij)$ at time $t_{I}$ it would be misleading to treat $I$ as a compound state and conclude that the underlying system remains in $I$ until the next transition $J$ is observed at $t_{J}$. At some time $t$ with $t_I \leq t \leq t_J$ the state of the coarse-grained system is described completely by knowing the last transition $I$ and the time $t - t_I$ that has passed since then. However, the same point in time on the reversed trajectory is described by knowing that $t_J - t$ has passed since the last transition $\widetilde{J}$. Thus, $\widetilde{J}$ replaces $I$ as the latest registered transition. Combining both effects allows us to formulate the time-reversal of a semi-Markov kernel $\psi_{I \to J}(t_{J} - t_{I})$ as
\begin{equation}
\widetilde{\psi}_{I\to J}(t_{J} - t_{I}) \equiv \psi_{\tilde{J}\to \tilde{I}}(t_{J} - t_{I})
\label{Eq:SMTimeRev}
\end{equation}
resulting in
\begin{equation}
    \mathcal{P}[\widetilde{\Gamma}|I_N,T] = \prod_{i=1}^{N-1}\psi_{\widetilde{I}_{i}\to \widetilde{I}_{i-1}}(t_i),
    \label{Eq:SMPWBack}
\end{equation}
for the conditioned path weight $\mathcal{P}[\widetilde{\Gamma}|I_N,T]$ of the time-reversed trajectory $\widetilde{\Gamma}$. Clearly, the time-reversal in \autoref{Eq:SMTimeRev} is identical to the time-reversal proposed in Ref. \cite{pedro2022} since the shift of inter-transition times discussed there is precisely the effect of kinetic hysteresis described above. Note that the modifications to the time-reversal operation of the semi-Markov process arise naturally, in accordance with the paradigm that time-reversal does not commute with coarse-graining of the form \autoref{Eq:Map} in general \cite{Hartich2020}.

In the common conception of semi-Markov processes, the direction-time independence criterion is a necessary condition to ensure time-reversal symmetry in equilibrium \cite{chari1994,Qian2007}. Remarkably, the semi-Markov process as introduced here breaks this condition in general. This apparent contradiction is resolved since the derivation of the direction-time independence relies crucially on the conventional time-reversal operation for semi-Markov processes, which does not apply here, as discussed above.

\subsection{Interpretation of the entropy estimators}

The entropy estimator $\braket{\hat{\sigma}}$ is established for unicyclic networks in \autoref{Eq:Sigma_Uni}. It is based on the microscopic fluctuation theorem in \autoref{eq:unicyclFluctTheo_Psi} valid for the ratio of waiting time distributions. The generalization of $\braket{\hat{\sigma}}$ for multicyclic networks with multiple observed links in \autoref{Eq:CEst}, which includes the estimator for a single observed link \autoref{Eq:BEst} as special case, relies on the same fluctuation theorem generalized to the multicyclic case. From the semi-Markov perspective, these fluctuation theorems can be interpreted as the consequence of an actual fluctuation theorem of the semi-Markov process. We define the semi-Markov entropy production rate $\sigma_{SM}$ as the limit
\begin{equation}
    \sigma_{\text{SM}} \equiv \lim_{T \to \infty} \frac{1}{T} \ln \frac{\mathcal{P}(\Gamma)}{\mathcal{P}(\widetilde{\Gamma})}
    \label{Eq:FlucT_SM}
,\end{equation}
which differs from the known expressions e.g. in \cite{andrieux2008, Esposito2008, maes2009} 
because of the modified time-reversal operation. Comparing \autoref{Eq:FlucT_SM} to \autoref{Eq:fluctTheoCC}, we conclude that $\sigma_{\text{SM}}$ in fact equals $\hat{\sigma}$, which was established as a thermodynamically consistent coarse-grained entropy production term in the previous sections. In hindsight, the fluctuation theorem in \autoref{eq:unicyclFluctTheo_Psi} can be derived from \autoref{Eq:FlucT_SM} by specifying to semi-Markov trajectories with only a single transition. The underlying Markov description does not enter explicitly anymore, instead it is incorporated implicitly by ensuring that $\sigma_{\text{SM}}$ is the correct physical entropy production. The affinity estimators derived in \autoref{sec:4} can also be seen as consequences of \autoref{Eq:FlucT_SM}, tracing back the entropy production to the level of contributing cycles.  

From the unifying semi-Markov perspective, we can give three complementary interpretations of the estimator $\braket{\hat{\sigma}}$. First, the derivation presented in Ref. \cite{pedro2022} relies on the information-theoretical identification of the expected entropy production of a stochastic process as a Kullback-Leibler divergence between the path weights of a forward and backward process \cite{roldan2010, roldan2012}. Second, contributions to the fluctuating quantity $\hat{\sigma}$ can be attributed to the completion of cycles in the underlying Markov network, which are partially observed for an external observer. Third, $\hat{\sigma} = \sigma_{\text{SM}}$ can be interpreted as the entropy production rate of a semi-Markov process with a particular time-reversal operation. Thermodynamic consistency of $\hat{\sigma}$ is then coupled to the applicability of the time-reversal operation, which has to be established from the underlying network. 

By interpreting $\hat{\sigma}$ as the entropy production $\sigma_{\text{SM}}$ of the equivalent semi-Markov process, the decomposition proposed in Ref. \cite{pedro2022} can be identified as a decomposition of $\langle\sigma_{\text{SM}}\rangle$ into the entropy production $\langle\sigma_{\text{EMC}}\rangle$ of the EMC and the remaining entropy production $\langle\sigma_{\text{WTD}}\rangle$ caused by the waiting times,
\begin{equation}
    \langle\sigma_{\text{SM}}\rangle = \langle\sigma_{\text{EMC}}\rangle + \langle\sigma_{\text{WTD}}\rangle.
    \label{Eq:S_Split}
\end{equation}
Up to a time conversion factor, $\langle\sigma_{\text{EMC}}\rangle$ is the mean entropy production of the EMC, which is given by
\begin{equation}
    \langle\sigma_{\text{EMC}}\rangle = \frac{1}{\braket{t}} \sum_{I,J}p_I^s p_{IJ}\ln\frac{p_{IJ}}{p_{\widetilde{J}\widetilde{I}}}
    \label{Eq:S_eMC},
\end{equation}
where $p_I^s$ is the steady state of the EMC as a discrete-time Markov chain. The factor $\braket{t}$, the average waiting time between two transitions, is needed because entropy production of a discrete-time Markov chain is naturally measured per step rather than per time. In terms of the application to observed links, $p_I^s$ quantifies the relative frequency of a particular transition $I$ in a long sequence of observed transitions as given by \autoref{Eq:SmEMC}. \add{Equivalently, \autoref{Eq:S_eMC} can be derived as the mean of}
\begin{equation}
    \sigma_{\text{EMC}} \equiv \lim_{T \to \infty} \frac{1}{T} \ln \frac{\mathcal{P}[\Gamma_{\text{EMC}}]}{\mathcal{P}[\widetilde{\Gamma}_{\text{EMC}}]},
\end{equation}
\add{defined on the level of single trajectories $\Gamma_{\text{EMC}}$ based on the arguments presented in appendix \ref{App:SM_Quantities}}. Note that \autoref{Eq:S_eMC} coincides with (49) of Ref. \cite{pedro2022}, dubbed there as transition sequence contribution to the entropy estimator. \add{Since the EMC emerges from integrating out the temporal resolution of the semi-Markov process, $\langle\sigma_{\text{EMC}}\rangle$ vanishes in situations with no observable net current. In other words, the contribution of a particular pair of transitions $I, J$ to $\sigma_{\text{EMC}}$ vanishes if and only if the net number of transitions $J$ after previous $I$ matches the number of transitions $\tilde{I}$ after previous $\tilde{J}$ on average, i.e., if $\mathcal{P}[\Gamma_{\text{EMC}}] = \mathcal{P}[\widetilde{\Gamma}_{\text{EMC}}]$.} 

\add{The condition of vanishing $\langle\sigma_{\text{EMC}}\rangle$ can also be related to the stalling conditions. In fact,} the entropy production associated with the embedded Markov chain coincides with the informed partial entropy estimator $\braket{\sigma_{IP}}$ formulated for the case of one accessible transition \cite{polettini2017, bisker2017}, i.e., 
\begin{equation}
    \braket{\sigma_{IP}} = \braket{\sigma_{\text{EMC}}}
    \label{Eq:S_ip_S_hat}
,\end{equation}
as proven in appendix \ref{Sec:supp:Prove_IP}. In particular, the force $F$ can be determined as
\begin{equation}
    \ln\frac{p_{I_+ I_+}}{p_{I_- I_-}} = \ln\frac{P(I_+|I_+)}{P(I_-|I_-)} = - F
    \label{Eq:F_IP_SM}
\end{equation}
by virtue of \autoref{Eq:F_IP} without referring to waiting times at all. This result is not surprising since both estimators measure the affinity $\mathcal{A}_{\mathcal{C}}$ of a single, averaged ''effective cycle`` either through the applied force $F$ or through the ratio $\ln P(+|+)/P(-|-)$. Without the time-resolution, the estimator $\braket{\hat{\sigma}}$ loses the ability to distinguish between longer or shorter hidden cycles. Thus, we can reformulate a conjecture proposed in Ref. \cite{pedro2022} that states that $\braket{\sigma_{\text{EMC}}}$ exceeds an analogous expression based on the TUR, $\langle\sigma_{TUR}\rangle$, since $\langle\sigma_{\text{EMC}}\rangle \geq \langle\sigma_{TUR}\rangle$ is equivalent to $\langle\sigma_{IP}\rangle \geq \langle\sigma_{TUR}\rangle$. As another consequence of \autoref{Eq:S_ip_S_hat}, the fluctuation theorem proven in \cite{bisker2017} for $\sigma_{IP}$, the fluctuating counterpart of the estimator $\braket{\sigma_{IP}}$, is related to its counterpart for the EMC, \autoref{Eq:S_eMC}.
 
The second expression in \autoref{Eq:S_Split}, $\langle\sigma_{\text{WTD}}\rangle$, can be deduced by transferring the splitting of the entropy production into contributions from the EMC and remaining contributions from the waiting times to individual semi-Markov kernels in the path weights. In more practical terms, a single semi-Markov kernel $\psi_{I\to J}(t)$ can be decomposed into
\begin{equation}
    \psi_{I\to J}(t) = p_{IJ}\cdot\psi(t|IJ)
    \label{Eq:WTD_Split},
\end{equation}
separating the contribution from the EMC from a conditional waiting time kernel $\psi(t|IJ) = \psi_{I\to J}(t)/p_{IJ}$. By decomposing all kernels in the path weights using \autoref{Eq:WTD_Split}, we can identify $\langle\sigma_{\text{WTD}}\rangle$ as a Kullback-Leibler divergence between the normalized probability densities $\psi(t|IJ)$ and their reverse, $\psi(t|\tilde{J}\tilde{I})$. Thus, the derivation in Ref. \cite{pedro2022} relates to factorizing out the EMC according to \autoref{Eq:WTD_Split} in the context of semi-Markov processes. \add{Using \autoref{eq:multiA}, we see that $\langle\sigma_{\text{WTD}}\rangle$ vanishes if and only if all $a_{IJ}(t)$ are constant in time. In particular, all $a_{IJ}(t)$ are constant in time if detailed balance is satisfied in the hidden subnetwork.}

The decomposition of the semi-Markov entropy production in \autoref{Eq:S_Split} clarifies additionally the relation between the estimator $\braket{\hat{\sigma}}$ and the entropy estimator $\braket{\sigma_{\text{KLD}}}$ introduced in \cite{martinez2019}, which is also decomposed in the form
\begin{equation}
    \braket{\sigma_{\text{KLD}}} = \braket{\sigma_{\text{aff}}} + \braket{\tilde{\sigma}_{\text{WTD}}}
.\end{equation}
Similar to \autoref{Eq:S_Split}, this decomposition into contributions from waiting time distributions and affinities is obtained by splitting off the EMC. The analogy is further strengthened by noting that 
\begin{equation}
    \braket{\sigma_{\text{aff}}} = \braket{\sigma_{IP}} = \braket{\sigma_{\text{EMC}}}
,\end{equation}
with the first equality proven in \cite{martinez2019}. Note that the respective embedded Markov chains are different \add{objects}, as $\braket{\sigma_{\text{aff}}}$ refers to a coarse-grained unicyclic three-state model, whereas $\braket{\sigma_{\text{EMC}}}$ only observes a single transition of this model. Nevertheless, the result is not entirely surprising in hindsight, since $\braket{\sigma_{\text{EMC}}}$ recovers the full entropy production of a unicyclic model by virtue of \autoref{eq:unicyclQuotient2}.

The difference between the estimators $\braket{\sigma_{\text{WTD}}}$ and $\braket{\tilde{\sigma}_{\text{WTD}}}$, or $\braket{\hat{\sigma}}$ and $\braket{\sigma_{\text{KLD}}}$, respectively, emerges from different rationales underlying the respective semi-Markov processes. Describing a physical system with a semi-Markov process is not sufficient to determine its entropy production uniquely, since the correct time-reversal operation needs to be discussed separately \cite{martinez2019, Hartich2021preprint,Bisker2022_Reply}. \add{In total, three different time-reversal operations for semi-Markov processes are implicitly used to define entropy estimators for partially accessible Markov networks.}

\begin{enumerate}
    \item Conventional time-reversal, $\tilde{\Gamma}(t) = \Gamma(T-t)$: In this case, physically consistent semi-Markov processes satisfy direction-time independence \cite{chari1994}, which causes $\braket{\sigma_{\text{WTD}}}$ to vanish \cite{Qian2007}. This time-reversal operation is applicable to particular settings of coarse-graining \cite{Qian2007, Ertel2021}. States do not change, i.e., are even under time-reversal.
    \item Modified time-reversal, introduced above: This operation includes the kinetic hysteresis effect introduced \cite{Hartich2020}, which is natural for coarse-graining based on milestoning \cite{Hartich2021}. In our case, semi-Markov states model transitions, which are odd under time-reversal.
    \item Time-reversal for second-order semi-Markov processes, introduced in \cite{martinez2019}: States in a second-order semi-Markov process are doublets containing the previous and current state by construction. Due to this memory effect, states are neither even nor odd under time-reversal.
\end{enumerate}

Any of these operations can be used to define an entropy via \autoref{Eq:FlucT_SM}. This entropy can always be split according to \autoref{Eq:WTD_Split}, where the resulting waiting time contributions are given by $0, \braket{\sigma_{\text{WTD}}}$ and $\braket{\tilde{\sigma}_{\text{WTD}}}$, respectively. In addition, any of the discussed operations are involutions, each giving rise to a dual dynamics for which an appropriate fluctuation theorem holds for the corresponding entropy production \cite{Seifert2012}. At this level, any non-vanishing entropy production quantifies a different mathematical notion of irreversibility, which becomes a thermodynamic quantity only if the time-reversal is known to be justified physically \cite{Hartich2020}.

\section{Conclusion}
\label{sec:7}

\subsection{Summary and Discussion}

In this paper, we have introduced an effective description for partially accessible Markov networks based on the observation of transitions along individual links and waiting times between successive observed transitions. The corresponding waiting time distributions yield an entropy estimator $\braket{\hat{\sigma}}$. The corresponding fluctuating counterpart $\hat{\sigma}$ additionally obeys a fluctuation theorem and was shown to have a natural interpretation as a semi-Markov entropy production. On a microscopic level, we have discussed with cycle fluctuation theorem arguments why observing one link suffices to recover the full entropy production in a unicyclic network. More generally, we have derived an operational criterion that indicates the absence of hidden cycles, which guarantees $\braket{\hat{\sigma}} = \braket{\sigma}$. 

If the hidden part of the network contains hidden cycles, we have shown that the estimator $\braket{\hat{\sigma}}$ yields a lower bound on the entropy production, which has been shown to improve on known estimation methods. Additionally, we have shown that the waiting time distributions contain information about topology and cycle affinities of the hidden network. To extract this information, we have derived exact results and estimation methods, whose quality has been assessed numerically. Both the entropy estimator and the affinity estimators are built upon the generalized microscopic cycle fluctuation theorem argument which is, as we have shown, the signature of a fluctuation theorem valid for an effective semi-Markov process. From the perspective of this semi-Markov process, we have unified extant entropy estimators by providing a mathematical interpretation.  

Different inference methods can be compared based on the required input data and the significance of their predictions. In the case of a single link, $\braket{\hat{\sigma}}$ relies on the measurement of statistical data contributing to a single current. While the amount of input data is comparable to methods based on the TUR, the predictions generally are much stronger, at least in the unicyclic case. While the TUR provides lower bounds on entropy production and cycle affinity in this case \cite{barato2015}, we recover exact values for both quantities even without access to the waiting times. When the waiting time distributions are available, exact cycle lengths can be deduced, which improves significantly on a known TUR-based trade-off relation between affinity and cycle length \cite{barato2015b, pietzonka2016}.

In terms of predictive significance, the entropy estimator is comparable to the method introduced in \cite{martinez2019} that is based on knowing a coarse-grained subnetwork, but it requires substantially less information. Calculating $\hat{\sigma}$ is possible without any knowledge about the underlying network beyond a single observed link. In particular, the issue of decimation schemes for coarse-graining is circumvented completely. Rather, the entropy estimator $\hat{\sigma}$ combines current measurements with information-theoretical notions via conditional counting, since our expectation on the next transition depends explicitly on the previous one \cite{roldan2010}. Thus, the sequence of transitions forms a Markov chain, which is identified as the EMC in the corresponding semi-Markov description. A mathematical discussion of semi-Markov processes allows us to clarify physically distinct categories of semi-Markov descriptions depending on the correct underlying time-reversal operation. Although different entropy-like quantities satisfy fluctuation theorems and provide a mathematical notion of irreversibility, the thermodynamically consistent entropy production must be identified by more fundamental means. If measuring the entropy production is feasible operationally, this knowledge can be used to decide which time-reversal operation recovers the correct entropy production. In this sense, identifying the correct time-reversal operation is a task of thermodynamic inference. 



\subsection{Perspectives}

The transition-based effective description for partially accessible Markov networks and the derived estimators for entropy and topology open a wide range of possible subsequent research topics. First of all, it will be promising to generalize the estimators for affinity and cycle length to networks with multiple observable links. Based on such a generalization, it would become possible to apply the estimators to a broader range of networks. The combined observation of different links would make it additionally possible to infer more information about the network because different affinities and cycle lengths would be accessible.  

\add{With the macroscopic limit of large, complex systems in mind, it is an obvious, albeit ambitious, challenge to transfer thermodynamic inference methods to Markov networks whose cycles outnumber the observed links by far. Conceptually, the ratio of waiting time distributions separates the time-resolved notion of irreversibility from other time-dependent effects entering a waiting time distribution. The estimation techniques for topology and affinity that are based on the short-time limit and hence short pathways infer local properties of Markov networks that may even be large. Passing from local to global methods would require a different approach. The dominant parts of the large-scale network structure might become manifest in patterns of particular transition sequences or waiting times in long trajectories. Splitting these into smaller snippets as proposed here is a first step towards a future study of self-correlations in a long trajectory to extract more complex structures.}

To gain more insight into the effective description from the established perspective of coarse-graining, one should investigate how existing coarse-graining strategies for observable states \cite{Knoch2015,Bo2017,Teza2020,shiraishi2015,polettini2017,Puglisi2010,Esposito2012,Bo2014,Mehl2012,Uhl2018,seiferth2020} are related to the approach introduced here. By combining these complementary approaches and by taking into account conclusions on milestoning \cite{Hartich2020,Hartich2021}, the concept of coarse-graining can potentially be generalized to a more fundamental level. From a practical perspective, we may ask how the method can be generalized to less ideal situations, e.g., if the observer cannot distinguish between different transitions or registers particular patterns or sequences of transitions only. This class of situations also includes the complementary problem when particular states can be observed rather than particular transitions, because observing the arrival in a state is equivalent to observing all transitions into this state without the ability to distinguish between them.

The potential of waiting time distributions and their role for inference schemes is certainly not exhausted by the results presented here. Combining the estimators for entropy production and network topology with existing numerical methods may increase the usefulness of the waiting time distributions in thermodynamic inference schemes. Fitting rates of the underlying Markov network to the recorded waiting time distribution \cite{Ehrich2021} or using minimization methods \cite{Skinner2021_1,Skinner2021_2} are promising tools to obtain tighter, more specialized bounds for the discussed estimators \add{or even to reconstruct the transition rates in a small network from sufficient data}. These methods will gain particular practical relevance since topological aspects of the underlying network can be deduced rather than have to be assumed.

Furthermore, even though the effective description has been introduced and discussed for observable transitions of a partially accessible Markov network in the NESS, it is, in principle, not limited to this setting. For example, the description could be applied beyond the steady state to analyze transient dynamics. Finally, it would be interesting to apply the approach to a Langevin dynamics to explore the adjustments needed for systems with continuous degrees of freedom.

\newpage

\appendix
\widetext
\setcounter{equation}{0}
\setcounter{section}{0}
\setcounter{subsection}{0}
\setcounter{figure}{0}
\setcounter{table}{0}
\makeatletter

\section{Waiting time distributions from path weights and trajectory snippets}
\label{sec:supp:absorbing}\label{App:PsiIdent}
\renewcommand{\theequation}{A\arabic{equation}}
\renewcommand{\thefigure}{A\arabic{figure}}

\subsection{Markovian path weights and master equation}

We consider the effective description of a given, only partially accessible system in which transitions are observed, e.g., the effective two-cycle network from the main text based on the observation of transitions between state 2 and state 3, shown in \autoref{fig:absorbingGraph}. We assume that there is an underlying, more fundamental network to which a discrete Markovian description from the perspective of stochastic thermodynamics, as described in detail in \cite{Seifert2012}, can be applied. For the effective description in \autoref{fig:absorbingGraph}(a), this full Markov network with two fundamental cycles is shown in \autoref{fig:absorbingGraph}(b). 

Transitions from state $k$ to state $l$ are governed by a transition rate $k_{kl}$, which is independent of the time already spent in the state $k$ due to the Markov property of the description. Thus, the waiting time distribution in a particular state must be memoryless and therefore exponentially distributed. In formulae,  the probability density for surviving in state $k$ until exactly time $t$ is given by $\Gamma_k \exp\left(-\Gamma_{k}t\right)$, where $\Gamma_{k}=\sum_{l}k_{kl}$ denotes the escape rate of state $k$. Given that state $k$ is exited, a transition to state $l$ is weighted with the transition rate and therefore happens with the transition probability $k_{kl}/\sum_l k_{kl}$.

Based on the discussed survival and transition probabilities, a path weight quantifying the probability of a trajectory $\zeta$ of the Markov network can be introduced. We assume that the network has $N$ states, is fully connected and that there are no unidirectional links, i.e., $k_{kl} > 0$ implies $k_{lk} > 0$. The path weight $\mathcal{P}[\zeta(t)]$ for a generic trajectory $\zeta(t)$ conditioned on the initial state $k_{0}$ at time $t = 0$ is given by
\begin{equation}
    \mathcal{P}[\zeta(t)|k_0,0] = \prod_{k=1}^{N}\exp\left(-\Gamma_{k}\tau_{k}\right)\prod_{(kl)}k_{kl}^{n_{kl}},
    \label{Eq:MarPW}
\end{equation}
where the second product runs over all possible transitions $(kl)$ in the network. The trajectory-dependent quantities $\tau_{k}$ and $n_{kl}$ denote the total time spent in state $k$ and the total number of transitions $(kl)$ in $\zeta(t)$, respectively. 

In principle, a trajectory-dependent observable can be obtained by a path integral over all trajectories $\zeta$, which in practice means summing over the number $L$ of possible jumps and integrating over all transition times $t_1, ..., t_L$. An important consequence is that the probability to observe $L$ jumps in a short trajectory $\zeta$ of length $\Delta t$ scales as $P(L \text{ jumps}) \sim \Delta t^L$ for $\Delta t \to 0$, since
\begin{equation}
    P(\zeta \text{ contains exactly } L \text{ jumps}|k_0,0) = \prod_{l=1}^L \left( \int _0 ^{\Delta t} dt_l \right) \mathcal{P}[\zeta(t)|k_0,0]\sim \Delta t ^L \left( 1 + \mathcal{O}(\Delta t) \right)
    \label{Eq:probLjumps}
\end{equation}
because the path weight as given in \autoref{Eq:MarPW} is of order $1$ in $\Delta t$. Thus, a first-order differential equation governing the time-evolution of $\zeta(t)$ can be derived by calculating the path weights for constant and one-jump trajectories, which are the only terms containing terms of first order in $\Delta t$. The resulting differential equation,
\begin{equation}
    \partial_t p_k(t) = \sum _{l \neq k} \left( p_l(t) k_{lk} - p_k(t) k_{kl}(t) \right)
    \label{Eq:masterEq}
\end{equation}
is known as the master equation and can be solved to obtain $p_k(t)$, the probability to find the system in state $k$ at time $t$. Since the master equation description \autoref{Eq:masterEq} is equivalent to the path weight description, solving the initial value problem for $p_k(0) = \delta_{k_0k}$ amounts to calculating
\begin{align}
    p_k(t) & = P(\zeta(t)=k|\zeta(0)=k_0) \label{Eq:propagator} \\
    & = \sum _{\zeta(t) = k} \mathcal{P}[\zeta(t)|k_0,0]
.\end{align}
The symbolic notation of a sum over paths will be used repeatedly in the following calculations.

\subsection{From fully accessible networks to partially accessible networks}

On a coarse-grained level of description, the trajectories of the network are only partially accessible. Thus, a complete analytical description by solving the master equation \autoref{Eq:masterEq} is generally impossible, because even the underlying fundamental network may be unknown. 

In the following, we assume that transitions along a single link connecting the Markov states $k$ and $l$ can be observed, but not the states themselves. This transition-based description coincides with the description proposed in \cite{pedro2022}. Adopting the notation from the main text, a transition along this link $k\to l$ and its reverse $l\to k$ are abbreviated as $I_{+} = (kl)$ and $I_{-} = (lk)$, respectively. Since the sequence of observed jumps and the waiting times in between are the only accessible information about the system in our effective description, a typical example of an observed effective trajectory $\Gamma$ may look like
\begin{equation}
    \Gamma ={} ? \to I_+ \to I_+ \to I_- \to I_+ \to \cdots \quad \text{at jump times } ?, T_0, T_1, T_2, T_3, ...
    \label{Eq:GammaTraj}
,\end{equation}
where $?$ represents the unknown transition of the system in the past prior to the first observed transition.

For simplicity, we assume from now on that the process starts and ends immediately after the observation of an observable transition, $I_1$ at time $T_0 = 0$, $I_{N+1}$ at time $T_N = T$, to address the core of our argumentation without worrying about non-time-extensive initial and final terms of the trajectory. Moreover, the scheme indicated in \autoref{Eq:GammaTraj} can be generalized to any number of observable links. We write $I_n = (k_n l_n)$ as the $n$-th observed transition between the underlying states $k_{n}$ and $l_{n}$, where we note that $l_n\neq k_{n+1}$ in general, as hidden dynamics cannot be excluded. Schematically, a coarse-grained trajectory $\Gamma$ takes the form
\begin{equation}
    \Gamma ={} I_1 \to I_2 \to I_3 \to \cdots \to  I_{N} \to I_{N+1} \quad \text{at jump times } T_0 = 0, T_1, T_2,  ..., T_{N-1}, T_N = T
    \label{Eq:GammaTraj_2}
.\end{equation}

Similar to the Markov case, the probability for $\Gamma$ can in principle be quantified by a path weight description. First of all, it is important to note that no memory effects ranging over multiple observed transitions need to be considered. The path weight for the future of the trajectory, i.e., the path weight for the trajectory after a transition $I_n$ is registered, is unaffected by the previous block $I_{n-1} \to I_n$ since knowing the transition $I_n = (k_nl_n)$ at $t_n$ implies knowing the state of the underlying markovian system immediately after $T_{n-1}$. Thus, the path weight can be split into parts belonging to observed transitions,
\begin{equation}
    \mathcal{P}[\Gamma(t)|I_1, 0] = P(I_{2}, T_1|I_1, 0) P(I_{3}, T_2|I_2, T_1) \cdots P(I_{N+1}, T_N|I_N, T_{N-1})
    \label{Eq:P_Split}
,\end{equation}
with $P(J, T_J|I, T_I)$ denoting the probability for observing transition $J$ at time $T_J$ if transition $I$ was observed at time $T_I$. Constituting the elementary building blocks in the coarse-grained picture, the objects $P(J, T_J|I, T_I)$ quantify the probability for observing $J$ after a given $I$ with waiting time $T_J - T_I$ in between. Thus, \autoref{Eq:P_Split} can also be written as
\begin{equation}
    \mathcal{P}[\Gamma(t)|I_1, 0] = \psi_{I_1 \to I_2}(t_1) \psi_{I_2 \to I_3}(t_2) \cdots \psi_{I_N \to I_{N+1}}(t_N),
    \label{Eq:P_Split_2}
\end{equation}
with $t_i = T_i - T_{i-1}$ and $t_1 = T_1$ in terms of the waiting time distribution
\begin{equation}
    \psi_{I \to J}(t) = P(J, t|I, 0)
    \label{Eq:PisPsi}
,\end{equation}
according to the definition of $\psi_{I \to J}(t)$ in \autoref{eq:setup:psiAsProb}. The waiting time distributions are normalized in the form
\begin{equation}
    \sum _J \int _0^\infty dt \; \psi_{I \to J}(t) = 1
,\end{equation}
whereas integrating out the time variable gives the marginal distribution
\begin{equation}
    p_{IJ} \equiv \int _0^\infty dt \; \psi_{I \to J}(t) = P(\text{next observed transition is } J| \text{last observed transition is } I)
    \label{eq:supp:psiInt}
.\end{equation}

\subsection{Effective absorbing dynamics}
\label{sec:absorbing}\label{App:Absorbing}

On a fundamental level, we are interested in how the path weights of the effective description \autoref{Eq:P_Split_2} and their elementary building blocks \autoref{Eq:PisPsi} are linked to the path weights \autoref{Eq:MarPW} of the corresponding microscopic trajectories of the full network. As a first step, we note that the way in which the effective trajectory $\Gamma$ was split carries over to a splitting on the fundamental level for the microscopic trajectory $\zeta$, because not only the coarse-grained but the entire microscopic state is known at the observed transition events. Symbolically, this can be denoted as
\begin{equation}
    \zeta\; \widehat{=}\; \gamma_{I_1 \to I_2}^{t_1}\; \to\; \gamma_{I_2 \to I_3}^{t_2}\; \to \; \cdots\; \to \; \gamma_{I_N \to I_{N+1}}^{t_N}
    \label{Eq:Snippet}
,\end{equation}
where $\gamma_{I \to J}^{t}$ is the snippet of the full trajectory between two subsequent observable transitions $I$ and $J$ with waiting time $t$ in between. This snippet starts in the destination state of $I$ and ends immediately after the transition event $J$ in the corresponding destination state. Since a given snippet is completed immediately after an observed transition $J$ is registered for the first time, each trajectory snippet can be interpreted as a trajectory of an effective Markovian absorbing dynamics defined on the full network obtained by removing all observed links. As soon as the original trajectory $\zeta$ completes an observed transition, the absorbing dynamics for $\gamma$ are terminated immediately. The corresponding first passage time is precisely the length of $\gamma$ in time and corresponds to the waiting time $t$ between two transitions in the effective description.

\begin{figure*}[bt]
    \begin{subfigure}[t]{0.22\textwidth}
      \centering
        \includegraphics[width=0.95\linewidth]{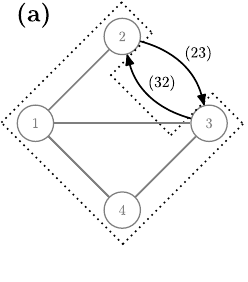}
    \end{subfigure}
    \hfill
    \begin{subfigure}[t]{0.22\textwidth}
      \centering
        \includegraphics[width=0.95\linewidth]{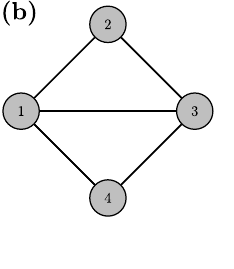}
    \end{subfigure}  
    \hfill
    \begin{subfigure}[t]{0.22\textwidth}
      \centering
         \includegraphics[width=0.95\linewidth]{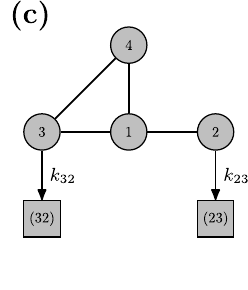}
    \end{subfigure} 
    \begin{subfigure}[t]{0.255\textwidth}
        \centering
        \vspace{-4.3cm}
            \includegraphics[width=\linewidth]{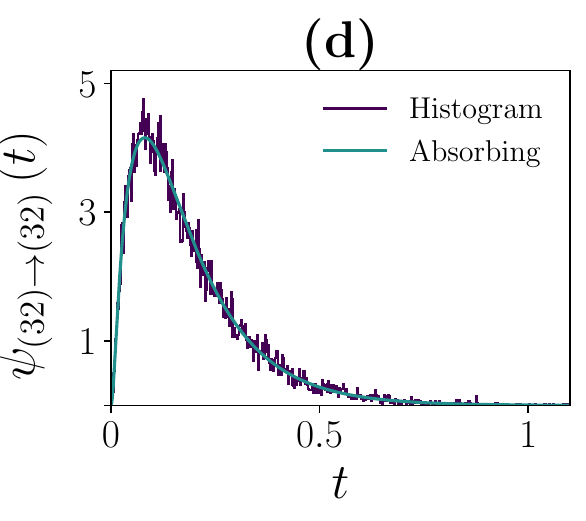}
    \end{subfigure}
    \caption[Absorbing Example]{Example for the analytical calculation of waiting time distributions based on effective absorbing dynamics. (a): Effective description for the partially accessible two-cycle network from the main text. Only transitions from state $2$ to state $3$ and in the reversed direction are observable. (b): Underlying Markov network. On the fundamental level of description, the network is markovian and transitions from state $k$ to state $l$ are governed by the transition rate $k_{kl}$. (c): Effective absorbing Markov network. Between two observed transitions, the system can be described with an absorbing master equation. This intermediate hidden dynamics is terminated by either a transition (32) or a transition (23). (d): Exemplary waiting time distribution derived from the numerical solution of the absorbing master equation for the effective dynamics and the corresponding distribution determined from a histogram of the waiting times within a trajectory of length $T = 10^{7}$ generated with a Gillespie simulation \cite{Gillespie_1977} of the network. The transition rates of the network were drawn randomly.}
\label{fig:absorbingGraph} 
\end{figure*}

Practically, the effective absorbing Markov network is obtained from the corresponding original network by treating all observable links as absorbing, i.e., redirecting the observed transitions into absorbing states. An example for such an effective absorbing Markov network is shown in \autoref{fig:absorbingGraph}(c), the corresponding absorbing network for the effective description of the two-cycle network in \autoref{fig:absorbingGraph}(a). The possible transitions along the observed link are represented by the states $(32)$ and $(23)$, which are absorbing states in the associated first passage problem. If the considered snippet begins with $(23)$ or $(32)$, the corresponding absorbing dynamics starts in $3$ or $2$, respectively.

The effective trajectory $\Gamma$ originates from a mapping of microscopic trajectories $\zeta \to \Gamma[\zeta]$ to the effective description of the system. The path weight of $\Gamma$ is obtained by summing over microscopic path weights
\begin{equation}
    \mathcal{P}[\Gamma(t)|I_1, 0] = \sum_{\zeta\in\Gamma} \mathcal{P}[\zeta(t)|l_1, 0]
    \label{Eq:PW2}
,\end{equation}
where $\mathcal{P}[\zeta|j_1, 0]$ is conditioned on $l_1$ at time $t=0$ for $I_1 = (k_1l_1)$. While integrating out the Markov path weight $\mathcal{P}[\zeta(t)]$ directly to obtain the coarse-grained path weight $\mathcal{P}[\Gamma(t)|I_1, 0]$ is not feasible in general, the decomposition of $\Gamma$ in \autoref{Eq:P_Split_2} and of $\zeta$ in \autoref{Eq:Snippet} reduces the problem to the level of elementary building blocks $\psi_{I \to J}(t)$ and $\gamma$, respectively. Thus, the decomposition \autoref{Eq:P_Split_2} can be combined with the summation in \autoref{Eq:PW2} to obtain
\begin{align}
    \mathcal{P}[\Gamma(t)|I_1, 0] = & \quad \quad \psi_{I_1 \to I_2}(t_1) \quad \; \; \quad \quad \quad \psi_{I_2 \to I_3}(t_2) 
    && \cdots \quad \; \; \quad \psi_{I_N \to I_{N+1}}(t_N) \\
    = & \sum_{\gamma \in \psi_{I_1 \to I_2}(t_1)} \mathcal{P}[\gamma|l_1, 0]
    \sum_{\gamma \in \psi_{I_2 \to I_3}(t_2)} \mathcal{P}[\gamma|l_2, 0]
    && \cdots \sum_{\gamma \in \psi_{I_N \to I_{N+1}}(t_N)} \mathcal{P}[\gamma|l_N, 0]
    \label{Eq:PW2_Split}
.\end{align}
The path weights $\mathcal{P}[\gamma|l_n, 0]$ of individual snippets $\gamma = \gamma_{I \to J}^t$ are seamlessly conditioned on the final state of their predecessor, since $I_n = (k_n l_n)$. The only type of summation that needs to be performed is the calculation of the waiting time distributions $\psi_{I\to J}(t)$ conditioned on $I=(kl)$ as introduced in \autoref{Eq:PisPsi} by integrating all possible $\gamma = \gamma_{I \to J}^t$, 
\begin{equation}
    \psi_{I\to J}(t) = \sum_{\gamma \in \psi_{I \to J}(t)}\mathcal{P}[\gamma_{I \to J}^t|l,0]
    \label{Eq:Psi_1}
.\end{equation}
This equation identifies the waiting time distributions of the effective description as summations over not observable trajectory snippets and proves therefore \autoref{eq:psiPathWeight} of the main text.

For $I=(kl)$ and $J=(mn)$, $\gamma_{I \to J}^t$ starts at $l$ and ends with a jump $(mn)$ exactly at time $t$. Since the system is in $m$ immediately before the jump at $t$, we can use the Markov property to calculate
\begin{align}
    \psi_{I\to J}(t) & = P(\text{jump }(mn) \text{ at time }t|l,0) \\
    & = P(\text{jump }(mn) \text{ at time }t|m,t)P(m,t|l,0) \\
    & = k_{mn} p_m(t)
    \label{Eq:Psi_Final}
,\end{align}
where \autoref{Eq:propagator} was used for the last equality. The result in \autoref{Eq:Psi_Final} makes it possible to calculate waiting time distributions analytically by solving the master equation of the effective absorbing dynamics defined on the hidden subnetwork. Note that this procedure is in principle equivalent to calculating the first passage time distributions for the associated first passage problem with the method introduced in \cite{sekimoto2021}. 

Conceptually, the reasoning used to derive \autoref{Eq:Psi_1} and therefore \autoref{Eq:Psi_Final} is identical to the reasoning used in \cite{pedro2022} to derive the inter-transition time densities. For both derivations, the partially accessible Markov network considered in the transition-based description is mapped to an effective first-passage time problem and the waiting time distributions are identified as the corresponding first passage time distributions. In our derivation, this mapping is motivated from an effective splitting emerging on the level of single trajectories, whereas in the derivation in \cite{pedro2022}, the mapping is deduced mathematically.

Operationally, the proposed calculation method for waiting time distributions differs from the method proposed in \cite{pedro2022}. Instead of carrying out the summation in \autoref{Eq:Psi_Final} explicitly, the waiting time distributions can be calculated from the solution of the effective absorbing master equation for different initial configurations using \autoref{Eq:Psi_1}. In addition, our calculation method is effective since collecting histogram data from a Gillespie simulation \cite{Gillespie_1977} is unnecessary to reconstruct the waiting time distributions, as they can be calculated directly.

To give an explicit example, the proposed method is used to calculate the waiting time distributions for the effective description of the two-cycle network in \autoref{fig:absorbingGraph}(a). Solving the corresponding effective absorbing master equation for fixed, randomly drawn transition rates results in four different waiting time distributions, one of them is shown in \autoref{fig:absorbingGraph}(d). Additionally, the figure shows how this waiting time distribution based on \autoref{Eq:Psi_Final} coincides with the corresponding waiting time distribution calculated from histogram data simulated with a Gillespie algorithm of the full network for long trajectories.

\section{Entropy estimator}
\label{sec:app:entropyestimator}\label{App:EntropyEst}
\renewcommand{\theequation}{B\arabic{equation}}
\renewcommand{\thefigure}{B\arabic{figure}}

\subsection{Coarse-grained and full entropy production}

Our effective description loses information about irreversibility and entropy production. From an abstract point of view, a well-defined many-to-one mapping of trajectories $\zeta \mapsto \Gamma[\zeta]$ of length $T$ suffices to bound the mean coarse-grained entropy production rate $\braket{\hat{\sigma}}$ against the physical entropy production rate $\braket{\sigma}$
\begin{equation}
    \braket{\hat{\sigma}} \equiv \frac{1}{T} \Braket{\ln \frac{\mathcal{P}[\Gamma]}{\mathcal{P}[\widetilde{\Gamma}]}} = \frac{1}{T} \sum _\Gamma \mathcal{P}[\Gamma] \ln \frac{\mathcal{P}[\Gamma]}{\mathcal{P}[\widetilde{\Gamma}]} \leq \braket{\sigma}
,\end{equation}
provided that $\Gamma \mapsto \widetilde{\Gamma}$ is the correct, physical time-reversal operation. Technically, the bound relies on the log-sum inequality, a standard tool in information theory \cite{ElemInfoTheo} stating
\begin{equation}
    \sum_i a_i \ln \frac{\sum_i a_i}{\sum_i b_i} \leq \sum_i a_i \ln \frac{a_i}{b_i}
    \label{eq:logsum}
,\end{equation}
for $a_i \geq 0$, $b_i \geq 0$. We apply this inequality in the form \cite{gomez-marin2008, seifert2018}
\begin{align}
    T \braket{\hat{\sigma}} & = \sum _{\zeta, \Gamma} \mathcal{P}[\Gamma|\zeta] \mathcal{P}[\zeta] \ln \frac{\sum_\zeta \mathcal{P}[\Gamma|\zeta] \mathcal{P}[\zeta]}{\sum_{\tilde{\zeta}} \mathcal{P}[\widetilde{\Gamma}|\tilde{\zeta}] \mathcal{P}[\tilde{\zeta}]} 
    \leq \sum _{\zeta, \Gamma} \mathcal{P}[\Gamma|\zeta] \mathcal{P}[\zeta] \ln \frac{\mathcal{P}[\Gamma|\zeta] \mathcal{P}[\zeta]}{ \mathcal{P}[\widetilde{\Gamma}|\tilde{\zeta}] \mathcal{P}[\tilde{\zeta}]} \\
    & = \sum _{\zeta, \Gamma} \mathcal{P}[\Gamma|\zeta] \mathcal{P}[\zeta] 
     \left( \ln \frac{\mathcal{P}[\Gamma|\zeta]}{ \mathcal{P}[\widetilde{\Gamma}|\tilde{\zeta}]}  + \ln \frac{\mathcal{P}[\zeta]}{\mathcal{P}[\tilde{\zeta}]} \right) = T \braket{\sigma}
.\end{align}
The last equality follows since $\mathcal{P}[\Gamma|\zeta] = 1$ is only satisfied if $\Gamma$ matches the correct effective trajectory $\Gamma[\zeta]$ and vanishes otherwise, $\mathcal{P}[\Gamma|\zeta] = 0$. Moreover, the equality requires that the first term in the sum vanishes, i.e. requires $\mathcal{P}[\widetilde{\Gamma}|\tilde{\zeta}] = 1$ when $\mathcal{P}[\Gamma|\zeta] = 1$. This condition defines the modified time-reversal operation $\Gamma \mapsto \widetilde{\Gamma}$ uniquely, since the correct $\widetilde{\Gamma}$ is identified as the trajectory obtained by using $\xi = \tilde{\zeta}$ in the mapping $\xi \mapsto \Gamma[\xi]$. In other words, we have to first time-reverse $\zeta$, which is then followed by the coarse-graining operation, as discussed in \cite{Hartich2020}.

\subsection{Time-reversal and conditional counting entropy estimator}
\label{sec:timeReversal}\label{App:TimeRev}

The previous section identifies the correct time-reversal operation $\Gamma \mapsto \widetilde{\Gamma}$ as the coarse-graining applied to the microscopic time reverse $\tilde{\zeta}(t) = \zeta(T - t)$. An effective trajectory $\Gamma$ consists of a series of transitions $I_n = (k_nl_n)$ at times $T_n$, which is schematically denoted
\begin{equation}
    \Gamma ={} (k_1l_1) \overset{t_1}{\to} (k_2l_2) \overset{t_2}{\to} (k_3l_3) \overset{t_3}{\to} \cdots \overset{t_{N-1}}{\to} (k_Nl_N) \overset{t_N}{\to} (k_{N+1}l_{N+1})
    \label{Eq:GammaTraj_3}
.\end{equation}
Compared to \autoref{Eq:GammaTraj_2}, the jumping times $T_i$ are replaced by the waiting times $t_i = T_i - T_{i-1}$ with $T_0 = 0$. Reversing the corresponding microscopic trajectory $\zeta$ in accordance with the previous discussion gives a well-defined effective trajectory of the form
\begin{align}
    \widetilde{\Gamma} & ={} (l_{N+1}k_{N+1}) \overset{t_N}{\to} (l_Nk_N) \overset{t_{N-1}}{\to} \cdots \overset{t_3}{\to} (l_3k_3) \overset{t_2}{\to} (l_2k_2) \overset{t_1}{\to} (l_1k_1)
    & = \widetilde{I}_{N+1} \overset{t_N}{\to} \widetilde{I}_N \overset{t_{N-1}}{\to} \cdots \overset{t_3}{\to} \widetilde{I}_3 \overset{t_2}{\to} \widetilde{I}_2 \overset{t_1}{\to} \widetilde{I}_1
    \label{Eq:GammaTraj_back}
,\end{align}
where we introduced the reversal operation on individual transitions $\widetilde{I}_{n} \equiv (l_nk_n)$ for $I_n = (k_nl_n)$. The reverse transition happens along the same link and is therefore also observable in the effective description by construction. The path weight for the backward trajectory \autoref{Eq:GammaTraj_back} can be decomposed into a product of single waiting time distribution objects as in \autoref{Eq:P_Split_2}, 
\begin{align}
    \mathcal{P}[\widetilde{\Gamma}(t)|I_{N+1}, 0] & = P[\tilde{I}_N, T_N - T_{N-1}|\tilde{I}_{N+1}, 0] P[\tilde{I}_{N-1}, T_N - T_{N-2}|\tilde{I}_N, T_N - T_{N-1}] \cdots P[\tilde{I}_1, T_N|\tilde{I}_2, T_N - T_1] \\
   & = \psi_{\tilde{I}_{N+1} \to \tilde{I}_N}(t_N) \; \psi_{\tilde{I}_N \to \tilde{I}_{N-1}}(t_{N-1}) \cdots \psi_{\tilde{I}_2 \to \tilde{I}_1}(t_1)
.\end{align}
After the proper time-reverse $\widetilde{\Gamma}$ is identified, the entropy production of a particular trajectory $\Gamma$ can be calculated explicitly as
\begin{align}
    T \hat{\sigma} = \ln \frac{P[\Gamma]}{P[\widetilde{\Gamma}]}  & =  \ln \frac{P(I_1)}{P(\tilde{I}_{N+1})} + \ln \frac{\mathcal{P}[\Gamma|I_1, 0]}{\mathcal{P}[\widetilde{\Gamma}|\tilde{I}_{N+1}, 0]} \\
    & = \ln \frac{P(I_1)}{P(\tilde{I}_{N+1})} + \sum _{j=1}^N \ln \frac{\psi_{I_j \to I_{j+1}}(t_j)}{\psi_{\widetilde{I}_{j+1} \to \widetilde{I}_j}(t_j)}  \\
    & = \ln \frac{P(I_1)}{P(\tilde{I}_{N+1})} + T \sum_{I,J} \int _0 ^\infty dt \, \nu_{J|I} (t) \ln \frac{\psi_{I \to J}(t)}{\psi_{\widetilde{J} \to \widetilde{I}}(t)}
    \label{Eq:TimeRev:Sigma}
,\end{align}
where the conditional counters $\nu_{J|I} (t)$ are introduced as
\begin{equation}
    \nu_{J|I} (t) \equiv \frac{1}{T} \sum _{j=1} ^N \delta(t_j - t) \delta_{I_{j+1}, J} \delta_{I_j, I}
.\end{equation}
In the limit $T \to \infty$, contributions from the initial and final state can be neglected, which yields the fluctuation theorem
\begin{align}
    \hat{\sigma} = \lim_{T \to \infty} \frac{1}{T} \ln \frac{P[\Gamma]}{P[\widetilde{\Gamma}]} = \sum_{I,J} \int _0 ^\infty dt \braket{\nu_{J|I} (t)} \ln \frac{\psi_{I \to J}(t)}{\psi_{\widetilde{J} \to \widetilde{I}}(t)}
\end{align}
and an explicit formula for the expected coarse-grained entropy production rate
\begin{align}
    \braket{\hat{\sigma}} = \sum_{I,J} \int _0 ^\infty dt \braket{\nu_{J|I} (t)} \ln \frac{\psi_{I \to J}(t)}{\psi_{\widetilde{J} \to \widetilde{I}}(t)}
    \label{eq:entropyCC}
.\end{align}



\subsection{Expectation values and entropy production for semi-Markov processes}
\label{App:SM_Quantities}
We will calculate the expectation value $\braket{\nu_{J|I} (t)}$ in \autoref{eq:entropyCC} using appropriate technique known for semi-Markov processes. Note that the transitions $I, J, ...$ are the ''states`` of the semi-Markov process and that the waiting time $t$ ''in state $I$`` is interpreted as the elapsed time since transition $I$. As defined in the main text, the conditional counter $\nu_{I \to J}(t)$ measures the number of transitions $J$ after a preceding transition $I$:

\begin{align}
    \nu_{I \to J}(t) \Delta t & = \frac{\text{No. of (}IJ\text{) jumps after waiting time} \in [t,t+\Delta t]}{T} \\
    \braket{\nu_{I \to J}(t)} & = P(\text{jump to }J \text{ after waiting time }t|I) \frac{\braket{\text{No. of jumps from }I}}{T} \\
    & = \psi_{I \to J}(t) \braket{n_I} = \psi_{I \to J}(t) \frac{p_I^s }{\braket{t}}
    \label{eq:suppErwNijt}
.\end{align}
In the last line, we have used
\begin{align}
    \braket{n_I} = \frac{\braket{\text{No. of jumps from }I}}{T} = \frac{\braket{\text{No. of jumps from }I}}{\braket{\text{Total No. of jumps}}} \cdot \frac{\braket{\text{Total No. of jumps}}}{T} = p_I^s \cdot \frac{1}{\braket{t}}
,\end{align}
where $\braket{t}$ is defined as the average waiting time between two semi-Markov transitions. The identification of the stationary distribution $p_I^s$ is based on elementary results for discrete-time Markov chains, as the number of visits of a particular state $I$ in a long trajectory $(I_1 I_2 ... I_N)$ divided by $N$ tends towards $p_I^s$ as $N \to \infty$. Note that, although this distribution is related to the stationary distribution of the semi Markov process itself, they are different even in the Markovian case \cite{Qian2007}. 

Since the $\psi_{I \to J}(t)$ are normalized by virtue of \autoref{eq:supp:psiInt}, we can integrate over $t$ to obtain the expected flux $\braket{n_{J|I}}$ from a semi-Markov state $I$ to $J$ as 
\begin{align}
    \braket{n_{J|I}} & = \int_0 ^\infty dt \, \braket{\nu_{I \to J}(t)} = p_{IJ} \frac{p_I^s}{\braket{t}} = \braket{n_I} p_{IJ} 
.\end{align}
The semi-Markov entropy production $\sigma_{SM}$ is defined by \autoref{Eq:FlucT_SM} as the probability ratio of forward and backward trajectory under the time-reversal operation $\Gamma \mapsto \tilde{\Gamma}$. Thus, the calculations of the previous \autoref{sec:timeReversal} starting from  \autoref{Eq:TimeRev:Sigma} actually apply to the semi-Markov entropy production $\sigma_{SM} = \hat{\sigma}$. Substituting \autoref{eq:suppErwNijt} into \autoref{eq:entropyCC}, we obtain
\begin{align}
    \braket{\sigma_{SM}} = \braket{\hat{\sigma}} = \sum_{I,J} \int _0 ^\infty dt \; \braket{n_I} \psi_{I \to J}(t) \ln \frac{\psi_{I \to J}(t)}{\psi_{\widetilde{J} \to \widetilde{I}}(t)}
\end{align}
for the semi-Markov entropy production. To put this expression into relation with the entropy production of the embedded Markov chain (eMC), we apply the log-sum inequality after using \autoref{eq:suppErwNijt} to obtain
\begin{align}
    \braket{\hat{\sigma}} \geq \frac{1}{\braket{t}} \sum_{I,J} p_I^s \left[\int _0 ^\infty dt \; \psi_{I \to J}(t) \right] \ln \frac{\int _0 ^\infty dt \; \psi_{I \to J}(t)}{\int _0 ^\infty dt \; \psi_{\widetilde{J} \to \widetilde{I}}(t)} = \frac{1}{\braket{t}} \sum_{I,J} p_I^s p_{IJ} \ln \frac{p_{IJ}}{p_{\widetilde{J}\widetilde{I}}} = \braket{\sigma_{EMC}}
    \label{eq:suppSigmaEMC}
\end{align}
in accordance with \autoref{Eq:S_eMC}.


\subsection{Comparison to informed partial entropy production}\label{Sec:supp:Prove_IP}

\subsubsection{Entropy estimators: embedded Markov chain vs. informed partial }

In this section, we prove that
\begin{equation}
    \braket{\sigma_{EMC}} = \braket{\sigma_{IP}}
\end{equation}
in the one-link case, which implies $\braket{\sigma_{IP}} \leq \braket{\hat{\sigma}}$ by virtue of \autoref{eq:suppSigmaEMC}. We will prove the case of one observable link between the Markov states $k$ and $l$, since the crucial relation \autoref{eq:supp:bigtheorem} and its proof can be generalized to multiple observed links following an analogous approach. For two states $+ = (kl)$ and $- = (lk)$, \autoref{eq:suppSigmaEMC} simplifies to 
\begin{align}
    \braket{\sigma_{EMC}} = \frac{1}{\braket{t}} \left( p_+^s p_{++} - p_-^s p_{--} \right) \ln \frac{p_{++}}{p_{--}} = \left( \braket{n_{+|+}} - \braket{n_{-|-}} \right) \ln \frac{\int _0 ^\infty dt\; \psi_{+ \to +}(t)}{\int _0 ^\infty dt\; \psi_{- \to -}(t)}
\end{align}
where \autoref{eq:suppErwNijt} and \autoref{eq:supp:psiInt} have been used for the second equality. It can be verified that for any sequence of transitions, i.e. any trajectory,  $n_{+|+} - n_{-|-}$ and $n_+ - n_-$ differ by at most $1$  due to terminal effects of the first and last transition that become negligible in the long-time limit. Thus, $\braket{n_{+|+} - n_{-|-}} = \braket{n_+ - n_-} = j$ can be used to express $\braket{\sigma_{EMC}}$ as 
\begin{equation}
    \braket{\sigma_{EMC}} = j \ln \frac{\int _0 ^\infty dt\; \psi_{+ \to +}(t)}{\int _0 ^\infty dt\; \psi_{- \to -}(t)}
    \label{eq:supp:sigmaEMC}
.\end{equation}
This expression is now in a form where it can be compared with the informed partial entropy production \cite{polettini2017, bisker2017}
\begin{align}
    \braket{\sigma_{IP}} = j \ln \frac{p^{st}_k k_{kl}}{p^{st}_l k_{lk}}
     \label{eq:supp:sigmaIP}
\end{align}
at the same link (connecting the Markov states $k$ and $l$). The stalling distribution $\textbf{p}^{st}$ is defined as the stationary state of the hidden subnetwork, where the link $k \to l$ is removed entirely. Thus, if the original process is generated by
\begin{align}
     \mathbf{L}_{ij} \equiv k_{ji} - \left( \sum_i k_{ji} \right) \delta_{ij}
,\end{align}
the stalling distribution satisfies $\mathbf{L}^{st} \mathbf{p}^{st} = 0$, with a modified generator $\mathbf{L}^{st}$ that is obtained from $\mathbf{L}$ by setting $k_{kl} = k_{lk} = 0$. The same stalling distribution $\textbf{p}^{st}$  can be accessed operationally if $k_{kl}$ and $k_{lk}$ depend on an external parameter $F$ via $k_{kl}(F)/k_{lk}(F) = \exp(F) k_{kl}(0)/k_{lk}(0)$. For $F=0$, the rates match their original value, $k_{kl}(0) = k_{kl}$ and $k_{lk}(0) = k_{lk}$, respectively. Then we can find the stalling distribution by sweeping over all values of $F$ until we find the value $F^{st}$ where the stationary current $j = p_k^{st} k_{kl}(F^{st}) - p^{st}_l k_{lk}(F^{st})$ vanishes. In a unicyclic process, this information is sufficient to infer the cycle affinity $\mathcal{A}$ \cite{bisker2017}, thus both $\braket{\sigma_{IP}}$ and $\braket{\hat{\sigma}}$ recover the full entropy production $\braket{\sigma} = j \mathcal{A}$ in this case. The relationship of the stalling distribution and the corresponding waiting time quotient can be rooted on a more general level. As we will prove, 
\begin{equation}
    \frac{p^{st}_k k_{kl}}{p^{st}_l k_{lk}} = \frac{\int _0 ^\infty dt\; \psi_{+ \to +}(t)}{\int _0 ^\infty dt\; \psi_{- \to -}(t)} 
    \label{eq:supp:bigtheorem}
\end{equation}
holds true in a general Markov network, which suffices to establish $\braket{\sigma_{IP}} = \braket{\sigma_{EMC}} \leq \braket{\hat{\sigma}}$ from \autoref{eq:supp:sigmaEMC} and \autoref{eq:supp:sigmaIP}. In particular,
\begin{align}
    \ln \frac{\braket{\int_0 ^\infty dt \, \nu_{+|+}(t)}/\braket{n_+}}{\braket{\int_0 ^\infty dt \, \nu_{-|-}(t)}/\braket{n_-}} = \frac{\int _0 ^\infty dt\; \psi_{+ \to +}(t)}{\int _0 ^\infty dt\; \psi_{- \to -}(t)} = \frac{p^{st}_k k_{kl}(0)}{p^{st}_l k_{lk}(0)} = \frac{k_{lk}(F) k_{kl}}{k_{kl}(F) k_{lk}} = - F
,\end{align}
where \autoref{eq:suppErwNijt} was used for the first equality. This result establishes \autoref{Eq:F_IP} in the main text.

\subsubsection{Proof of equation (\ref{eq:supp:bigtheorem})}
To establish \autoref{eq:supp:bigtheorem}, we consider the absorbing problem from Appendix \ref{sec:supp:absorbing} with two absorbing states ''$+$`` and  ''$-$``, which indicate an observed forward jump $k \to l$ and backward jump $l \to k$, respectively. In the notation of \autoref{sec:supp:absorbing}, we have ''$+$``$=(kl)$ and ''$-$``$=(lk)$. The absorbing problem can be seen as a discrete-time Markov chain with transition probabilities $p_{i\to j}$ given by
\begin{align}
    p_{i \to j} = \begin{cases} \frac{k_{kl}}{\sum_j k_{kj}} & \text{if } i=k, j=+ \\
        \frac{k_{lk}}{\sum_j k_{lj}} & \text{if } i=l, j=- \\
        0 & \text{if } i=\pm \\
        0 & \text{if } i=k, j=l \text{ or } i=l, j=k\\
        \frac{k_{ij}}{\sum_j k_{ij}} & \text{else}
        \end{cases}
    \label{eq:supp:pijdef}
.\end{align} Here, $i,j \in \{ 1, ..., N, +, - \}$ can be any one of the $N$ Markov states or the absorbing states $\pm$.

Since the states $+$ and $-$ are absorbing, the Markov chain will end in one of these states almost surely. The probability that the last state is $+$ or $-$ given the current state $j$ will be denoted $P(\pm|j)$. These probabilities are related to the waiting time distributions via
\begin{align}
    \int _0^\infty dt \; \psi_{+ \to +}(t) & = P(+|(kl)) = P(+|l) 
    \label{eq:supp:psiToP1}\\
    \int _0^\infty dt \; \psi_{- \to -}(t) & = P(-|(lk)) = P(-|k)
    \label{eq:supp:psiToP2}
\end{align}
by virtue of \autoref{eq:supp:psiInt}. To calculate $P(+|j)$ and $P(-|j) = 1 - P(+|j)$, we observe that these probabilities have to satisfy a linear system of equations
\begin{equation}
    P(+|i) = \sum _j p_{i\to j} P(+|j)
    \label{eq:supp:lgs1}
,\end{equation}
with analogous equations for $P(-|j)$. It is sufficient to let $1 \leq j \leq N$, since $P(+|+) = P(-|-) = 1 - P(+|-) = 1 - P(-|+) = 1$. The system in \autoref{eq:supp:lgs1} can be recast in matrix form by adjusting the generator matrix $\mathbf{L}$ slightly. We define $\widetilde{\mathbf{L}}$ by setting the $(k,l)$-th and $(l,k)$-th element of $\mathbf{L}$ to zero, i.e.,
\begin{align}
     \widetilde{\mathbf{L}}_{ij} \equiv \begin{cases} 0 & i=k, j=l \text{ or } i=l, j=k \\
     \mathbf{L}_{ij} & \text{else}
     \end{cases}
    \label{eq:supp:ltilde}
.\end{align}
Note that $\widetilde{\mathbf{L}}$ is not the generator of a Markov process, as it is not column-stochastic. After substituting \eqref{eq:supp:pijdef} into \autoref{eq:supp:lgs1} and multiplying both sides with $\sum_j k_{ij}$, we rearrange terms to obtain
\begin{align}
    \widetilde{\mathbf{L}}^T \mathbf{p}_+ & = - k_{kl} \mathbf{e}_k \label{eq:supp:pplus1} \\
    \widetilde{\mathbf{L}}^T  \mathbf{p}_- & = - k_{lk} \mathbf{e}_l \label{eq:supp:pminus1}
,\end{align}
with the vectors $\mathbf{p}_\pm \equiv \sum_i P(\pm|i) \mathbf{e}_i$. Applying Cramer's rule, we can express the inverse matrix $\widetilde{\mathbf{L}}^{-1}$ as
\begin{align}
    \left(\widetilde{\mathbf{L}}^T\right)^{-1}_{ij} = \frac{\det \left( ( \widetilde{\mathbf{L}}^T)_{(j|i)}\right)}{\det \widetilde{\mathbf{L}}^T} (-1)^{i+j}  = \frac{\det \widetilde{\mathbf{L}}_{(i|j)}}{\det \widetilde{\mathbf{L}}} (-1)^{i+j}
.\end{align}
Here, $\mathbf{L}_{(j|i)}$ denotes the matrix obtained by removing the $j$-th row and the $i$-th column from $\widetilde{\mathbf{L}}$. By multiplying \autoref{eq:supp:pplus1} and \autoref{eq:supp:pminus1} with $\widetilde{\mathbf{L}}^{-1}$, we obtain
\begin{align}
    P(+|l) & = - k_{kl} \widetilde{\mathbf{L}}^{-1}_{lk} = - k_{kl} \frac{\det \widetilde{\mathbf{L}}_{(l|k)}}{\det \widetilde{\mathbf{L}}} (-1)^{l+k} \label{eq:supp:pplus2} \\
    P(-|k) & = - k_{lk} \widetilde{\mathbf{L}}^{-1}_{kl} = - k_{lk} \frac{\det \widetilde{\mathbf{L}}_{(k|l)}}{\det \widetilde{\mathbf{L}}} (-1)^{k+l} \label{eq:supp:pminus2}
\end{align}
for the $k$-th row of \autoref{eq:supp:pplus1} and the $n$-th row of \autoref{eq:supp:pminus1}, respectively. Dividing these equations, we obtain
\begin{equation}
    \frac{P(+|l)}{P(-|k)} = \frac{k_{kl}}{k_{lk}} \frac{\det \widetilde{\mathbf{L}}_{(l|k)}}{\det \widetilde{\mathbf{L}}_{(k|l)}}
    \label{eq:supp:PtoL}
.\end{equation}
This equality can be compared to a relation connecting the stalling distribution $\mathbf{p}^{st}$ to the generator $\mathbf{L}^{st}$,
\begin{equation}
    \frac{\det \mathbf{L}_{(l|k)}^{st}}{\det \mathbf{L}_{(k|l)}^{st}} = \frac{p_k^{st}}{p_l^{st}}
    \label{eq:supp:biskerP}
,\end{equation}
which is established in Appendix B of \cite{bisker2017}. Comparing the matrices $\mathbf{L}^{st}$ and $\widetilde{\mathbf{L}}$ (cf. \autoref{eq:supp:ltilde}), we observe that they only differ in the $(k,k)$-th and the $(l,l)$-th element, since the absorbing states affect the escape rates in state $k$ and $l$ only. In particular, the minors 
\begin{align}
    \det \mathbf{L}_{(l|k)}^{st} & = \det \widetilde{\mathbf{L}}_{(l|k)} \\
    \det \mathbf{L}_{(k|l)}^{st} & = \det \widetilde{\mathbf{L}}_{(k|l)}
\end{align}
coincide, since $\mathbf{L}_{(l|k)}^{st} = \widetilde{\mathbf{L}}_{(l|k)}$ and similarly for interchanged $l \leftrightarrow k$. By using these identities and the equations \eqref{eq:supp:psiToP1}, \eqref{eq:supp:psiToP2}, \eqref{eq:supp:PtoL} and \eqref{eq:supp:biskerP}, the claim in \autoref{eq:supp:bigtheorem} follows from
\begin{equation}
    \frac{\int _0 ^\infty dt\; \psi_{+ \to +}(t)}{\int _0 ^\infty dt\; \psi_{- \to -}(t)} =  \frac{P(+|l)}{P(-|k)} = \frac{p^{st}_k k_{kl}}{p^{st}_l k_{lk}}
.\end{equation}

\section{Cycle length estimation from waiting time distributions}\label{App:L_Est}
\renewcommand{\theequation}{C\arabic{equation}}
\renewcommand{\thefigure}{C\arabic{figure}}

As shown in \autoref{sec:supp:absorbing}, the waiting time distributions of an observed system can be determined by solving the absorbing master equation \autoref{Eq:masterEq} for the corresponding escape problem. Microscopic trajectories associated with a waiting time distribution of the form $\psi_{I\to J}(t)$ for transitions $I=(kl)$, $J=(mn)$ starting in $l$ must reach $m$. As implied by \autoref{Eq:probLjumps}, the short-time behavior of the corresponding path weight encodes the number of transitions needed for the completion of this path. 

\subsection{Length of the shortest connection}\label{App:L_Est_N1}
The length $N_1$ of the shortest connection between $l$ and $m$ in the hidden subnetwork, i.e. the minimal number of hidden transitions, is a lower bound on the number of transitions of any trajectory from $l$ to $m$. By virtue of \autoref{Eq:probLjumps}, we have the scaling
\begin{equation}
    p_m(t) \sim t^{N_1}
\end{equation}
for the solution $p_m(t)$ of the corresponding absorbing master equation for $p_l(0) = 1$ as $t \to 0$. With \autoref{Eq:Psi_Final}, this carries over to
\begin{equation}
    \psi_{I\to J}(t) \sim t^{N_{1}} 
\label{Eq:PsiScale}
.\end{equation}
To extract $N_{1}$ from $\psi_{I\to J}(t)$, we take the logarithm of \autoref{Eq:PsiScale}, which results in
\begin{equation}
    \ln\psi_{I\to J}(t)\sim N_{1}\ln t
    \label{Eq:PsiScale_N11}
.\end{equation}
Thus, the derivative with respect to $\ln t$ scales as
\begin{equation}
    \frac{d\left(\ln\psi_{I\to J}(t)\right)}{d\left(\ln t\right)}\sim N_{1},
    \label{Eq:PsiScale_N12}
\end{equation}
for small $t$, which ultimately gives
\begin{equation}
     \add{\lim_{t \to 0} \left( t \frac{d}{dt} \ln \psi_{I_{\pm}\to I_ {\pm}}(t) \right) = N_{1}}
    \label{Eq:PsiScale_N13}
\end{equation}
in the limit $t \to 0$. \add{Note that the logarithmic derivative in \autoref{Eq:PsiScale_N12} has been written out explicitly in \autoref{Eq:PsiScale_N13} to increase readability.} Based on \autoref{Eq:PsiScale_N13}, the length $N_1$ of the shortest connection between $l$ and $m$ can be estimated from the short time limit of a given waiting time distribution for transitions $I=(kl)$ and $J=(mn)$.

\subsection{Length of the second shortest connection}\label{App:L_Est_N2}
The second shortest connection between $I=(kl)$ and $J=(mn)$ can be estimated similarly by considering the short time limit of the ratio of waiting time distributions. To this end, we calculate the number of links in the second shortest connection between the Markov states $l$ and $m$ from a time scale separation. The logarithmic ratio of waiting time distributions $a_{I\to J}(t)$ is defined as
\begin{equation}
    a_{I\to J}(t) = \ln \frac{\sum_{\gamma_{I \to J}^t} \mathcal{P}[\gamma_{I \to J}^t|I]}{\sum_{\gamma_{\widetilde{J} \to \widetilde{I}}^t} \mathcal{P}[\gamma_{\widetilde{J} \to \widetilde{I}}^t|\widetilde{J}]},
    \label{Eq:N2_1}
\end{equation}
where $\gamma_{I \to J}^t$ denotes the snippet of the underlying trajectory starting with $I$ and ending with $J$ after time $t$ with corresponding conditioned path weight $\mathcal{P}[\gamma_{I \to J}^t|I]$. As defined in the main text and in \autoref{sec:timeReversal}, the tilde on a capital letter reverses the corresponding transition, e.g.  $\widetilde{I} = (lk)$ for $I=(kl)$.

In a first step, we divide the sets of all possible snippets $\gamma_{I \to J}^t$ and their time reversed counterparts $\gamma_{\widetilde{J} \to \widetilde{I}}^t$ into two subsets corresponding to their number of jumps. Let $n_\gamma$ denote the number of hidden transitions between the states $l$ and $m$ in the snippet  $\gamma_{\widetilde{J} \to \widetilde{I}}^t$. By definition of $N_1$, we have $n_\gamma \geq N_1$, which allows us to write
\begin{equation}
    a_{I\to J}(t) = \ln \frac{\sum_{N_1 \leq n_\gamma < N_2} \mathcal{P}[\gamma_{I \to J}^{t}|I] + \sum_{n_\gamma \geq N_2} \mathcal{P}[\gamma_{I \to J}^{t}|I]}{\sum_{N_1 \leq n_\gamma < N_2} \mathcal{P}[\gamma_{\widetilde{J} \to \widetilde{I}}^{t}|\widetilde{J}] + \sum_{n_\gamma \geq N_2} \mathcal{P}[\gamma_{\widetilde{J} \to \widetilde{I}}^{t}|\widetilde{J}]}.
    \label{Eq:N2_2}
\end{equation}
for a natural number $N_2 > N_1$ which, at this point, is arbitrary. By using \autoref{Eq:probLjumps}, we note that for small $t$ the two sums scale as
\begin{align}
    \sum_{N_1 \leq n_\gamma < N_2} \mathcal{P}[\gamma_{I \to J}^{t}|I] & \simeq t^{N_1} \left[1 + \mathcal{O}(t^{N_1+1}) \right], \label{Eq:N2_scaling1} \\
   \qquad \sum_{n_\gamma \geq N_2} \mathcal{P}[\gamma_{I \to J}^{t}|I] & \simeq t^{N_2} \left[1 + \mathcal{O}(t^{N_2+1}) \right]
   \label{Eq:N2_scaling2}
.\end{align}
Since reversing a path does not change the number of jumps, the sums in the denominator satisfy analogous scaling laws. We will now choose $N_2$ as the largest number for which 
\begin{equation}
   a_0 = \ln \frac{\sum_{N_1 \leq n_\gamma < N_2} \mathcal{P}[\gamma_{I \to J}^{t}|I]}{\sum_{N_1 \leq n_\gamma < N_2} \mathcal{P}[\gamma_{\widetilde{J} \to \widetilde{I}}^{t}|\widetilde{J}]}
   \label{Eq:N2_3}
\end{equation}
is time-independent. Such a number exists if the shortest connection of length $N_1$ is unique, since in this case $N_1 + 1$ is a valid choice. The reason is that if only paths containing $N_1$ transitions between $l$ and $m$ are allowed, only one possible connection remains, i.e. $a_0$ is time-independent because the network becomes unicyclic effectively. This argument remains valid as long as $N_2$ is smaller than or equal to the length of the second shortest connection. As soon as $n_\gamma$ is large enough to allow for paths along the second connection, $a(t)$ becomes time-dependent if the resulting hidden cycle has nonvanishing affinity.  

Thus, we proceed knowing that $N_2$ as defined in \autoref{Eq:N2_3} is the length of the second shortest connection. Substituting this equation into \autoref{Eq:N2_2}, we obtain 
\begin{equation}
    a_{I\to J}(t) = a_0 + \ln \frac{1 + \sum_{n_\gamma \geq N_2} \mathcal{P}[\gamma_{I \to J}^{t}|I]/\sum_{N_1 \leq n_\gamma < N_2} \mathcal{P}[\gamma_{I \to J}^{t}|I]}{1 + \sum_{n_\gamma \geq N_2} \mathcal{P}[\gamma_{\widetilde{J} \to \widetilde{I}}^{t}|\widetilde{J}]/\sum_{N_1 \leq n_\gamma < N_2} \mathcal{P}[\gamma_{\widetilde{J} \to \widetilde{I}}^{t}|\widetilde{J}]}
    \label{Eq:N2_4}
.\end{equation}
After rearranging terms, we use \autoref{Eq:N2_scaling1} and \autoref{Eq:N2_scaling2} to extract the lowest order correction in the numerator,
\begin{equation}
    \ln \left(1 + \sum_{n_\gamma \geq N_2} \mathcal{P}[\gamma_{I \to J}^{t}|I]/\sum_{N_1 \leq n_\gamma < N_2} \mathcal{P}[\gamma_{I \to J}^{t}|I] \right) \sim t^{N_2-N_1}
,\end{equation}
and similarly for the denominator.  After rearranging terms in \autoref{Eq:N2_4},
\begin{equation}
    \Delta a_{I\to J}(t) \equiv a_{I\to J}(t) - a_0 \sim t^{N_2-N_1}
,\end{equation}
taking the limit $t \to 0$ yields the result
\begin{equation}
    \add{\lim_{t \to 0} \left( t \frac{d}{dt} \ln \left| \Delta a_{I\to J}(t)\right| \right) = N_{2} - N_{1}}
.\end{equation}

\section{Multiple observed links: Full entropy production if hidden network is trivial}
\label{sec:app:multiProof}\label{App:Multi_FullEnt}
\renewcommand{\theequation}{D\arabic{equation}}
\renewcommand{\thefigure}{D\arabic{figure}}

We will now prove the second part of rule 1 in \autoref{sec:5b}., which asserts that the full entropy production is recovered in the long-time limit $T \to \infty$, i.e.,
\begin{equation}
    \braket{\sigma} = \braket{\hat{\sigma}}
\end{equation}
if the hidden subnetwork is topologically trivial. We start by recalling the total medium entropy production $s_{m}$ of a Markov trajectory $\zeta = (n_0 n_1 ... n_M)$ and its corresponding fluctuation theorem \cite{Seifert2012}
\begin{equation}
    s_m[\zeta] \equiv \sum_{j=1}^M \ln \frac{k_{n_{j-1}n_j}}{k_{n_jn_{j-1}}} = \ln \frac{\mathcal{P}[\zeta(t)|n_0]}{\mathcal{P}[\tilde{\zeta}(t)|n_M]}
.\end{equation}
In the long-time limit $T \to \infty$, it is known that
\begin{equation}
    \braket{\sigma} = \frac{1}{T} \braket{s_m}
    \label{eq:supp:medEntIsSigma}
,\end{equation}
since the difference between $\sigma$ and $s_m/T$ is a stochastic entropy term that is not extensive in time. Using the splitting introduced in \autoref{Eq:Snippet}, the medium entropy production can be decomposed similarly, which is written here somewhat symbolically as
\begin{equation}
    s_{m}[\zeta] = \sum_{j = 1}^N \sigma_{m}[\gamma_{I_j \to I_{j+1}}^{t{_j}}] + \sum_{j = 1}^{N+1} \ln \frac{k_{k_jl_j}}{k_{l_jk_j}}
    \label{eq:supp:medEnt}
.\end{equation}
The second term contains the entropy production due to the visible transitions $I_j = (k_jl_j)$ ($j = 1, ..., N+1$) with respective transition rates $k_{I_j}$ and their reverse $k_{\tilde{I}_j}$. In contrast, the first term includes the exact medium entropy production of a trajectory snippet $\gamma_{I_j \to I_{j+1}}^{t{_j}}$ in the hidden subnetwork, i.e. without including the contributions from the visible initial and final transitions $I_j$ and $I_{j+1}$, respectively.

Since the hidden subnetwork satisfies detailed balance, as it does not contain any hidden cycles, it is possible to identify a well-defined potential function $F(i)$ on the Markov states $i$ satisfying
\begin{equation}
    F(i) - F(j) = \ln \frac{k_{ij}}{k_{ji}}
\end{equation}
for neighboring states $i$, $j$ in the hidden subnetwork. With this potential function, the stochastic entropy production of a trajectory in the hidden subnetwork becomes path-independent: A trajectory $\gamma$ starting and ending in states $l$ and $k$ respectively satisfies $\sigma_{m}[\gamma] = F(l) - F(k)$. In particular, we can use the fluctuation theorem property \autoref{eq:supp:medEnt} of $\sigma_m$ to identify
\begin{equation}
    F(l_j) - F(k_{j+1}) = s_{m}[\gamma_{I_j \to I_{j+1}}^{t{_j}}] = \ln \frac{\mathcal{P}[\gamma_{I_j \to I_{j+1}}^{t{_j}}|I_j]}{\mathcal{P}[\gamma_{\tilde{I}_{j+1} \to \tilde{I}_j}^{t{_j}}|\tilde{I}_{j+1}]} - \ln \frac{k_{k_{j+1}l_{j+1}}}{k_{l_jk_j}}
    \label{eq:supp:medEnt2}
\end{equation}
Due to the existence of the potential, the result is independent of the particular sequence of transitions in $\gamma_{I_j \to I_{j+1}}^{t{_j}}$. Note that the second term is needed because the observed transitions that terminate a trajectory snippet were not accounted for in $\sigma_{m}$, which it is defined in the hidden subnetwork.

Since the hidden subnetwork is topologically trivial, we can evaluate the ratio of path weights in \autoref{eq:supp:medEnt2} using \autoref{eq:multiFluctTheo}. We obtain
\begin{equation}
    a_{I_jI_{j+1}} = \ln \frac{\mathcal{P}[\gamma_{I_j \to I_{j+1}}^{t{_j}}|I_j]}{\mathcal{P}[\gamma_{\tilde{I}_{j+1} \to \tilde{I}_j}^{t{_j}}|\tilde{I}_{j+1}]} = s_{m}[\gamma_{I_j \to I_{j+1}}^{t{_j}}] + \ln \frac{k_{k_{j+1}l_{j+1}}}{k_{l_jk_j}}
\end{equation}
for the trajectory snippets between two observed jumps. Reassembling these individual contributions via \autoref{eq:supp:medEnt} yields
\begin{equation}
    s_{m}[\zeta] = \sum_{j=1}^N a_{I_jI_{j+1}} + \ln \frac{k_{k_1l_1}}{k_{l_{N+1}k_{N+1}}}
    \label{eq:supp:medEnt3}
.\end{equation}
The first term recovers the value of $\hat{\sigma}$ precisely, because \autoref{Eq:CEst} reduces to
\begin{equation}
    \hat{\sigma} = \sum_{IJ} a_{IJ} n_{J|I} = \frac{1}{T} \sum_{j=1}^N a_{I_jI_{j+1}}
    \label{eq:supp:medEnt4}
\end{equation}
since $a_{IJ}(t) = a_{IJ}$ is time-independent. To calculate the corresponding averages, we note that the second term of \autoref{eq:supp:medEnt3} is negligible in the limit $T \to \infty$. By combining \autoref{eq:supp:medEntIsSigma}, \autoref{eq:supp:medEnt3} and \autoref{eq:supp:medEnt4}, we conclude
\begin{equation}
    \braket{\sigma} = \frac{1}{T} \braket{s_{m}} = \braket{\hat{\sigma}}
.\end{equation}
This last equality verifies the claim in rule 1 in \autoref{sec:5b}.

\section{Model parameters, simulation and counter example}
\label{sec:app:modelparameters}\label{App:Para}
\renewcommand{\theequation}{E\arabic{equation}}
\renewcommand{\thefigure}{E\arabic{figure}}

\subsection{Unicyclic network for the illustration of the coarse-graining scheme in Figure 1}

The unicyclic network in Fig. 1 of the main text includes four different states with different links in between. The transition rates for the network are shown in \autoref{Tab:Rates_Uni}. The waiting time distributions shown in Fig. 1 (d) of the main text can be derived from the solution of the absorbing master equation for the effective absorbing Markov network defined by Fig. 1 (b) of the main text with the method from \autoref{sec:supp:absorbing}. For the waiting time distributions shown in Fig. 1 (d) of the main text, the corresponding absorbing master equation for the transition rates in \autoref{Tab:Rates_Uni} has been solved numerically.

\begin{table}[H]
    \caption{Transition rates for the unicyclic network in Fig. 1}
    \centering
    \begin{tabular}{l|l|l|l}
        \hline
        State 1 & State 2 & State 3 & State 4\\
        \hline
        \hline
        $k_{12} = 1.2$& $k_{21} = 1.7$& $k_{31} = 1.0$& $k_{41} = 1.9$\\
        $k_{13} = 3.6$& $k_{23} = 1.0$& $k_{32} = 1.0$& $k_{41} = 1.9$\\
        $k_{14} = 2.0$& & $k_{34} = 1.8$& \\
         \hline
    \end{tabular}
\label{Tab:Rates_Uni}
\end{table}

\subsection{Multicyclic network for the illustration of the affinity bounds in Figure 2}

The multicyclic network in Fig. 2 of the main text includes seven different states with different links in between. The transition rates for the network leading to the affinities given in the caption of Fig. 2 of the main text are shown in \autoref{Tab:Rates_Multi1}. This specific choice of transition rates results in different time scales for the transitions through the individual cycles including the observable link. Thus, no cycle is preferred and the existence of all cycles can be inferred from the successive maxima and minima of the corresponding $a(t)$.

$a_{(71)\to(71)}(t)$ in Fig. 2 (c) and (d) of the main text is by definition given by the logarithm of the fraction of the corresponding waiting time distributions. As discussed in \autoref{sec:supp:absorbing}, these waiting time distributions can be derived from the solution of the absorbing master equation for the effective absorbing Markov network defined by Fig. 2 (b) of the main text. For $a_{(71)\to(71)}(t)$ shown Fig. 2 (c) and (d) of the main text, the corresponding absorbing master equation for the transition rates in \autoref{Tab:Rates_Multi1} has been solved numerically.

\begin{table}[H]
    \caption{Transition rates for the multicyclic network in Fig. 2}
    \centering
    \begin{tabular}{l|l|l|l|l|l|l}
        \hline
        State 1 & State 2 & State 3 & State 4 & State 5 & State 6 & State 7\\
        \hline
        \hline
        $k_{12} = 1.0$& $k_{21} = 1.0$& $k_{31} = 1.0$& $k_{43} = 5.0$& $k_{54} = 3.0$& $k_{65} = 1.0$& $k_{71} = 1.0$\\
        $k_{13} = 3.0$& $k_{23} = 1.0$& $k_{32} = 8.0$& $k_{45} = 3.0$& $k_{56} = 4.0$& $k_{67} = 30.0$& $k_{72} = 0.1$\\
        $k_{17} = 1.0$& $k_{27} = 0.1$& $k_{34} = 40.0$&  & $k_{57} = 0.5$&  & $k_{75} = 50.0$\\
         &  &  &  &  &  & $k_{76} = 2.0$\\
         \hline
    \end{tabular}
\label{Tab:Rates_Multi1}
\end{table}

\subsection{Multicyclic network for the quality factors of the affinity bounds in Figure 3}

The multicyclic network from Fig. 2 of the main text is also used in Fig. 3 of the main text to illustrate the quality of the derived affinity bounds. For each realization of the network, the transition rates are drawn randomly according to the distributions given in \autoref{Tab:Rates_Multi2}. All transition rates are distributed uniformly. The different choices for the definition intervals lead on the one hand to a bias in the affinity $\mathcal{A}_{\mathcal{C}_{0}}$ towards positive values and increase on the other hand the maximal possible value of $\mathcal{A}_{\mathcal{C}_{0}}$ aiming at a broad range of possible affinities for the scatter plots of the quality factors in Fig. 3 (b), (c) and (d) of the main text. As a consequence of this bias, it is more likely to draw high positive affinities rather than high negative ones. Since the bounds for the maximal affinities are tighter for single, high positive affinities, the quality factors $\mathcal{Q}_{+}$ in Fig. 3 (c) are centered around higher values in comparison to the quality factors $\mathcal{Q}_{-}$ in Fig. 3 (d).  

To calculate the quality factors for the two different classes of network configurations shown in Fig. 3 (b), (c) and (d) of the main text, the maximum and minimum of $a_{(71)\to(71)}(t)$ are needed for all randomly drawn realization of the multicyclic network. For a given realization, $a_{(71)\to(71)}(t)$ can be calculated from the waiting time distributions that can be derived by using the absorbing master equation method from \autoref{sec:supp:absorbing}. For each realization corresponding to one quality factor shown in Fig. 3 (b), (c) and (d) of the main text, the absorbing master equation for the drawn transition rates has been solved numerically, $a_{(71)\to(71)}(t)$ has been determined and the values of the maximum and minimum of $a_{(71)\to(71)}(t)$ have been used to calculate the corresponding quality factor according to the definitions given in the caption of Fig. 3 of the main text. 

\begin{table}[H]
    \caption{Distributions of the random transition rates for the multicyclic network in Fig. 3. $\Theta(a,b)$ denotes a uniform distribution defined between $a$ and $b$.}
    \centering
    \begin{tabular}{l|l|l|l|l|l|l}
        \hline
        State 1 & State 2 & State 3 & State 4 & State 5 & State 6 & State 7\\
        \hline
        \hline
        $k_{12}: \Theta(0.01,80)$& $k_{21}: \Theta(0.01,2)$& $k_{31}: \Theta(0.01,2)$& $k_{43}: \Theta(0.01,2)$& $k_{54}: \Theta(0.01,2)$& $k_{65}: \Theta(0.01,2)$& $k_{71}: \Theta(0.01,80)$\\
        $k_{13}: \Theta(0.01,20)$& $k_{23}: \Theta(0.01,80)$& $k_{32}: \Theta(0.01,2)$& $k_{45}: \Theta(0.01,20)$& $k_{56}: \Theta(0.01,20)$& $k_{67}: \Theta(0.01,20)$& $k_{72}: \Theta(0.01,2)$\\
        $k_{17}: \Theta(0.01,2)$& $k_{27}: \Theta(0.01,80)$& $k_{34}: \Theta(0.01,20)$&  & $k_{57}: \Theta(0.01,20)$&  & $k_{75}: \Theta(0.01,20)$\\
         &  &  &  &  &  & $k_{76}: \Theta(0.01,20)$\\
         \hline
    \end{tabular}
\label{Tab:Rates_Multi2}
\end{table}

\subsection{Two-cycle network for topology and the entropy estimators in Figure 4}

The two-cycle network in Fig. 4 of the main text includes four different states and five links in total. Since the network is comparatively compact yet nontrivial, it was also used as an example network for multiple extant entropy estimators \cite{polettini2017,bisker2017,martinez2019,Ehrich2021}. To compare the introduced entropy estimator $\hat{\sigma}$ to existing results, the rates given in \autoref{Tab:Rates_TS} are chosen according to the rates in \cite{martinez2019}. The force parameter $F$ applied to the observable link is related to the transition rates of the observed link via
\begin{equation}
    - 2 F = \ln \frac{k_{23}}{k_{32}} - \ln \frac{2}{3}
.\end{equation}
This parameter is introduced in \cite{martinez2019} to tune the observed net current $j = n_+ - n_-$. To estimate the cycle lengths as discussed in Fig. 4 (b) and (c) of the main text, the waiting time distributions for the observable transition are needed to calculate $\ln\psi_{(32)\to (32)}(t)$ and $a_{(32)\to (32)}(t)$. As for the networks in Fig. 1 and Fig. 2, these distributions can be derived by using the absorbing master equation method from \autoref{sec:supp:absorbing}. For the estimations of the cycle lengths shown in Fig. 4 (b) and (c), the corresponding absorbing master equation for the transition rates in \autoref{Tab:Rates_Multi1} with $F = \ln 3$ has been solved numerically.

To calculate the introduced entropy estimator $\hat{\sigma}$ for a given trajectory of the two cycle network, $a(t)$ and the conditioned transition counters $\nu_{+|+}(t)$ and $\nu_{-|-}(t)$ of the observed link are needed. $a(t)$ can be calculated independently of $\nu_{+|+}(t)$ and $\nu_{-|-}(t)$ from the corresponding waiting time distributions derived with the absorbing master equation method from \autoref{sec:supp:absorbing}. A trajectory resulting from the observation of the two cycle network can be generated by simulating the full underlying Markov network with the Gillespie algorithm \cite{Gillespie_1977}. For the entropy estimator in Fig. 3 (c), $a(t)$ has been calculated from the waiting time distributions resulting from the numerical solution of the corresponding absorbing master equation. The conditioned transition counters have been calculated by counting the corresponding transitions in a simulated Gillespie trajectory of length $T = 10^{7}$, weighting the calculated number of transitions with the corresponding $a(t)$ evaluated at the registered waiting time leads to $\langle\hat{\sigma}\rangle$. The mean entropy production of the Markov network have been directly calculated in the Gillespie simulation.  

\begin{table}[H]
    \caption{Transition rates for the two cycle network in \autoref{Fig:Diamond}. $F$ is a dimensionless force applied to the observable link between state 2 and state 3.}
    \centering
    \begin{tabular}{l|l|l|l}
        \hline
        State 1 & State 2 & State 3 & State 4\\
        \hline
        \hline
        $k_{12} = 1.0$& $k_{21} = 8.0$& $k_{31} = 0.2$& $k_{41} = 75.0$\\
        $k_{13} = 35.0$& $k_{23} = 3.0\cdot\exp\left(-F\right)$& $k_{32} = 2.0\cdot\exp\left(F\right)$& $k_{43} = 2.0$\\
        $k_{14} = 0.7$&  & $k_{34} = 50.0$&  \\
         \hline
    \end{tabular}
\label{Tab:Rates_TS}
\end{table}

\subsection{A model with an invisible cycle}
\label{sec:counterexample}

In this section, we present an explicit Markov network that has a cycle with nonvanishing affinity in the hidden subnetwork, although all $a_{IJ}(t)$ are time-independent. We use the network of \autoref{Fig:Diamond} with the choice of rates given by \autoref{Tab:Rates_counterex}. Instead of observing the transitions between state $2$ and $3$, we observe $I_+ = (13)$ and $I_- = (31)$.

\begin{table}[H]
    \caption{Transition rates for the counterexample. The network is given by \autoref{Fig:Diamond} with the specified configuration of rates.}
    \centering
    \begin{tabular}{l|l|l|l}
        \hline
        State 1 & State 2 & State 3 & State 4\\
        \hline
        \hline
        $k_{12} = x$  & $k_{21} = 1.0$& $k_{31} = 1.0$& $k_{41} = 1.0$\\
        $k_{13} = 1.0$& $k_{23} = 1.0$& $k_{32} = 1.0$& $k_{43} = 1.0$\\
        $k_{14} = 1.0$&  & $k_{34} = 1.0$&  \\
         \hline
    \end{tabular}
\label{Tab:Rates_counterex}
\end{table}

This network has the remarkable property that lumping state $2$ and state $4$ into a single compound state $H$ results in a Markov network. The new transition rates are given by 
\begin{align}
    k_{1H} & = k_{12} + k_{14} = x + 1 \\
    k_{3H} & = k_{32} + k_{34} = 2 \\
    k_{H1} & = k_{21} = k_{41} = 1 \\
    k_{H3} & = k_{23} = k_{43} = 1
.\end{align}
The ratio $a(t)$ remains unaffected by this procedure. In particular, since the model is now reduced to a unicyclic network with cycle $\mathcal{C} = (13H1)$, $a(t)$ is constant in time, taking the value $a(t) = \mathcal{A}_{\mathcal{C}} = \ln [2/(a+1)]$. In contrast, the cycle affinity of the true hidden cycle $\mathcal{C'} = (12341)$ in the hidden subnetwork is given by $\mathcal{A}_{\mathcal{C}'} = \ln a$. For $a \neq 1$ this model is an example for which $a(t)$ is constant in time despite the presence of a cycle with nonvanishing affinity in the hidden subnetwork.

\twocolumngrid

\newpage

\end{document}